\documentclass{aastex}
\usepackage{spr-astr-addons}             
\usepackage{url}
\usepackage{lscape, morefloats, graphicx, float}   
\RequirePackage{color}                   

\begin{document}

\title{Metallicity Calibration and Photometric Parallax Estimation: I.{\em UBV} photometry}
\slugcomment{Not to appear in Nonlearned J., 45.}
\shorttitle{Metallicity Calibration and Photometric Parallax Estimation}
\shortauthors{S. Tun{\c c}el G\"u{\c c}tekin, Bilir, S., Karaali, S., Ak, S., T. Ak, Z. F. Bostanc\i}

\author{S. Tun{\c c}el G\"u{\c c}tekin \altaffilmark{1}}
\altaffiltext{1}{Istanbul University, Graduate School of Science and Engineering, 
Department of Astronomy and Space Sciences, 34116, Beyaz\i t, Istanbul, Turkey\\
\email{sabihaguctekin@gmail.com}}

\author{S. Bilir \altaffilmark{2}} 
\altaffiltext{2}{Istanbul University, Faculty of Science, Department 
of Astronomy and Space Sciences, 34119 University, Istanbul, Turkey\\}
\and
\author{S. Karaali \altaffilmark{2}} 
\altaffiltext{2}{Istanbul University, Faculty of Science, Department 
of Astronomy and Space Sciences, 34119 University, Istanbul, Turkey\\}
\and
\author{S. Ak \altaffilmark{2}}
\altaffiltext{2}{Istanbul University, Faculty of Science, Department 
of Astronomy and Space Sciences, 34119 University, Istanbul, Turkey\\}
\and
\author{T. Ak\altaffilmark{2}} 
\altaffiltext{2}{Istanbul University, Faculty of Science, Department 
of Astronomy and Space Sciences, 34119 University, Istanbul, Turkey\\}
\and
\author{Z. F. Bostanc\i\altaffilmark{2}} 
\altaffiltext{2}{Istanbul University, Faculty of Science, Department 
of Astronomy and Space Sciences, 34119 University, Istanbul, Turkey\\}

\begin{abstract} 
We present metallicity and photometric parallax calibrations for the F and G type dwarfs with photometric, astrometric and spectroscopic data. The sample consists of 168 dwarf stars covering the colour, iron abundance and absolute magnitude intervals $0.30<(B-V)_0<0.68$ mag, $-2.0<[Fe/H]<0.4$ dex and $3.4<M_V<6.0$ mag, respectively. The means and standard deviations of the metallicity and absolute magnitude residuals are small, i.e. $\langle\Delta[Fe/H]_{res}\rangle=0$ and $\sigma=0.134$ dex, and $\langle\Delta (M_V)_{res}\rangle=0$ and $\sigma=0.174$ mag, respectively, which indicate accurate metallicity and photometric parallax estimations.

\end{abstract}

\keywords{stars: abundances, stars: metallicity calibration, stars: distance}

\section{Introduction}
Stellar metallicity and kinematics are two important means to understand the formation and evolution of our Galaxy. Metallicity of a star can be determined spectroscopically or photometrically. One needs more accurate analyses for the  first procedure  which can be carried out only for nearby stars, while the second one, the photometric procedure, can be applied to stars at large distances as well. The second procedure has a long history. We are indebted to \citet{Roman55} who discovered the correlation between the weakness of the spectral lines and  the ultra-violet (UV) excess, $\delta(U-B)$, of the stars.  

It has been a custom to use the $(U-B)\times(B-V)$ two-colour diagram of the Hyades cluster to estimate the $\delta(U-B)$ colour excesses, i.e. $\delta(U-B)$ of a star is the difference between the UV colours  of the given star and the Hyades' star with the same $B-V$ colour. \citet{Schwarzschild55}, \citet{Sandage59} and \citet{Wallerstein62} interpreted the UV excess with the ``blanketing model'', i.e. the UV excesses of the stars, with the same metal abundance but different $B-V$ colours are different. This is important for the red stars where the blanketing line is parallel to the intrinsic Hyades line and hence $\delta(U-B)$ is guiotined. \citet{Sandage69} used a procedure to normalize the UV excesses, as explained in the following. He separated 112 high proper motion stars into 16 sub-samples with mean colours $B-V$=0.35, 0.40, 0.45, ..., 1.10, and compared the UV excess for each colour, $\delta(U-B)$, with the one for $(B-V)_0=0.6$, $\delta_{0.6}(U-B)$, which is the maximum UV excess for a given star. Then, he adopted the ratio $\delta(U-B)_{0.6}/\delta(U-B)$ as a normalized factor for the UV excess for the corresponding $B-V$ colour.

\citet{Carney79} normalized the UV excesses of 101 dwarfs using the procedure of \citet{Sandage69} and calibrated them to the iron abundance $[Fe/H]$. \citet{Karaali03} improved this calibration using a different procedure and 88 dwarf stars. These calibrations provide iron abundances in the $UBV$ photometry. Other studies offered calibration in different photometries, including $UBV$, \citep[i.e.][]{Walraven60, Stromgren66, Cameron85, Laird88, Buser90, Trefzger95}. Two recent metallicity calibrations which provide iron abundances with $UBV$ photometry are those of \citet{Karatas06} and \citet{Karaali11}.

The distance of a star is an important parameter in investigation of the kinematic structure of our Galaxy which can be determined via trigonometric or photometric parallaxes. Trigonometric parallax procedure can be applied only to nearby stars. {\it Hipparcos} \citep{ESA97, vanLeeuwen07} is the main source for this parallax. For stars at large distances, only photometric parallax is available, which is the combination of the apparent and absolute magnitude of a star in question. Absolute magnitude estimation is another problem for the researchers. The widely used procedure for absolute magnitude estimation is based on the offset from a standard main-sequence, where Hyades is generally used for this purpose \citep{Laird88, Nissen91, Karaali03, Karatas06}. 

The colour-absolute magnitude of a specific cluster can also be used for absolute magnitude estimation of stars in a given population, such as thin and thick discs, and halo. The cluster is chosen such that its metallicity is compatible with the mean metallicity of the population in question. Details for this procedure can be found in \citet{Phleps00}, \citet{Chen01}, and \citet{Siegel02}.

In this study, we present two calibrations; one for the iron abundance $[Fe/H]$, and another one for the absolute magnitude offset $\Delta M_V$, in terms of reduced UV excess, $\delta_{0.6}(U-B)$. We used the reduced {\it Hipparcos} astrometric data \citep{vanLeeuwen07} for the first time and improved both the  metallicity and absolute magnitude calibrations. We organized the paper as follows. The data are presented in Section 2. The procedure is given in Section 3, and Section 4 is devoted to Summary and Discussion. 

\section{Data}

The data used in our study is a combination of the star samples given in four spectroscopic studies, i.e. \citet{Bensby14}, \citet{Nissen10}, \citet{Reddy06}, and \citet{Venn04}. We separated the F-G dwarf stars in these studies with the temperature and surface gravity constraints of \citet{Cox00}, i.e. $5310<T_{eff}(K)<7300$, and $\log g>4$ (cgs) and reduced the multiple observations to a single one. The number of stars taken from the mentioned studies in the order given above which supply the constraints of \citet{Cox00} are 263, 83, 51, and 32, respectively. After excluding the binary and variable stars from the sample, as well as the multiple observations, the total star sample reduced to 168. The iron abundances of these stars were also taken from the mentioned four spectroscopic studies. We used the reduced {\it Hipparcos} parallaxes \citep{vanLeeuwen07} and estimated the distances of the sample stars using the following equation:

\begin{equation}
d({\rm pc})=1000/\pi ({\rm mas}).\\
\end{equation}

The $V$ apparent magnitudes and, $U-B$ and $B-V$ colours of the sample stars are provided from the catalogue of \citet{Mermilliod97} and they are de-reddened by the procedure explained in the following. We adopted the total absorption for the model, $A_{\infty}$, \citep{Schlafly11} and estimated the total absorption for the distance of the star, $A_d$, by the following equation of \citet{Bahcall80}:

\begin{equation}
A_{d}(b)=A_{\infty}(b)\Biggl[1-\exp\Biggl(\frac{-\mid
d~\sin(b)\mid}{H}\Biggr)\Biggr],
\end{equation} 
where $b$ and $d$ are the Galactic latitude and distance of the star, respectively, and $H$ is the scale-height of the dust \citep[$H=125$ pc;][]{Marshall06}. Then, the colour excess of the star, $E_d(B-V)$, could be estimated by the following equation of \citet{Cardelli89} and was used in the evaluation of the colour excess $E_d(U-B)$ given by \citet{Garcia88}:

\begin{equation}
E_d(B-V)=A_d(b)/3.1,
\end{equation}

\begin{equation}
E_d(U-B)=0.72\times E_d(B-V)+0.05\times E_d(B-V)^2.    
\end{equation}
Finally, the de-reddened $V_0$ magnitude and colour indices are:
\begin{eqnarray}
V_0= V-3.1\times E_d(B-V),\\ \nonumber
(B-V)_0=(B-V)-E_d(B-V),\\ \nonumber
(U-B)_0=(U-B)-E_d(U-B).\\ \nonumber
\end{eqnarray}
Thus, we estimated the absolute magnitudes of the stars using the necessary parameters in the well known Pogson equation, i.e.
\begin{equation}
V_0-M_V=5\times \log d-5.
\end{equation}

\begin{figure}[t]
\begin{center}
\includegraphics[scale=0.40, angle=0]{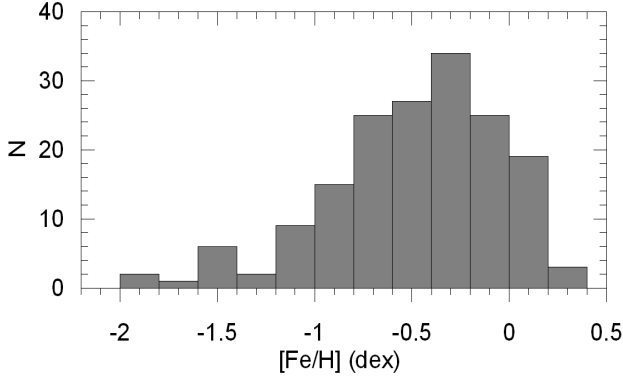}
\caption[] {Metallicity distribution of the sample stars.}
\end{center}
\end{figure}

Data for the 168 sample stars are given in Table 1. They cover the colour range $0.30<(B-V)_0<0.68$ mag. The distributions of the iron abundance, relative parallax errors and the absolute magnitudes are given in the Figs. 1-3. The mode value of the metallicity histogram in Fig. 1 corresponds to -0.3 dex, while the iron abundances cover the range $-2<[Fe/H]<+0.4$ dex. However, the number of stars with $[Fe/H]<-1.2$ and $[Fe/H]>0.2$ dex are only 14. The number of stars with relative parallax error $\sigma_{\pi}/\pi\leq 0.1$ is 141, which comprise about 84\% of the complete sample. The mode value of absolute magnitudes of the sample stars is calculated to be $M_V=4.5$ mag. The absolute magnitudes in Fig. 3 range from 3.4 to 6.0 mag. The data for the two-colour diagram, $(U-B)_0\times(B-V)_0$, and $M_V\times(B-V)_0$ colour-absolute magnitude diagram for the Hyades cluster were taken from \citet{Sandage69} and \citet{Karaali03}, respectively, and then they were fitted to fifth degree polynomial equations as follows:

{\small
\begin{eqnarray}
(U-B)_0=2.791(0.702)-20.567(5.558)\times(B-V)_0\\ \nonumber
+58.355(16.920)\times(B-V)_0^2-81.837(24.836)\times(B-V)_0^3\\ \nonumber
+59.22(17.631)\times(B-V)_0^4-17.115(4.857)\times(B-V)_0^5.\\\nonumber
\end{eqnarray} }
and 
{\footnotesize
\begin{eqnarray}
M_V=5.661(3.185)-36.582(22.392)\times(B-V)_0\\ \nonumber
+133.228(58.043)\times(B-V)_0^2-188.632(72.402)\times(B-V)_0^3\\ \nonumber
+122.703(43.577)\times(B-V)_0^4 -29.961(10.156)\times(B-V)_0^5.\\ \nonumber
\end{eqnarray}}
The numbers in parenthesis indicate the error values of the related coefficients. The correlation coefficient and standard deviation of the Eq. (7) and Eq. (8) are $R^2=0.999$ and $\sigma=0.007$ mag and $R^2=0.989$ and $\sigma=0.164$ mag, respectively.  

\begin{figure}[t]
\begin{center}
\includegraphics[scale=0.40, angle=0]{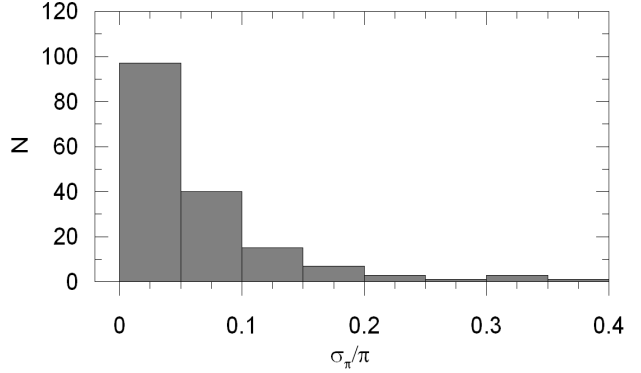}
\caption[] {Distribution of the relative parallax errors for the sample stars.} 
\end{center}
\end {figure}

\begin{figure}[H]
\begin{center}
\includegraphics[scale=0.40, angle=0]{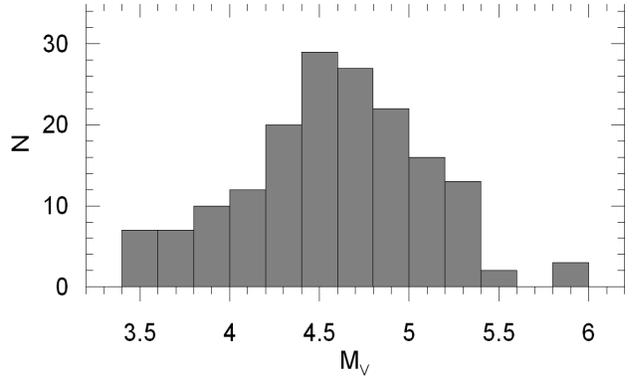}
\caption[] {Absolute magnitude distribution for the sample stars.}
\end{center}
\end{figure}

\section{Procedure}
\subsection{Metallicity Calibration}
We used the same procedure in \citet{Karaali05} and \citet{Karatas06} for the metallicity calibration. However, our data are different than the cited studies. As stated in Section 2, our metallicities (the iron abundances) are provided from recent studies \citep{Bensby14, Nissen10, Reddy06, Venn04}. Hence, we expect more accurate iron abundances. The same case holds for the absolute magnitudes and distances which are estimated via the re-reduced {\it Hipparcos} parallaxes \citep{vanLeeuwen07} and hence they are more accurate than the ones published formerly \citep{ESA97}. We fitted the guillotine factors of \citet{Sandage69} in sixteen $B-V$ colours to a sixth degree polynomial as in the following which could be used to evaluate a guillotine factor (normalized factor) for each $B-V$ colour:

{\footnotesize
\begin{eqnarray}
f=\delta(U-B)_{0.6}/\delta(U-B)=8.175-83.963\times(B-V)_0\\ \nonumber
+409.970\times(B-V)_0^2-1018.242\times(B-V)_0^3\\ \nonumber
+1341.239\times(B-V)_0^4-889.844\times(B-V)_0^5\\ \nonumber
+234.468\times(B-V)_0^6.\\ \nonumber
\end{eqnarray} 
}

\begin{figure}[h]
\begin{center}
\includegraphics[scale=0.4, angle=0]{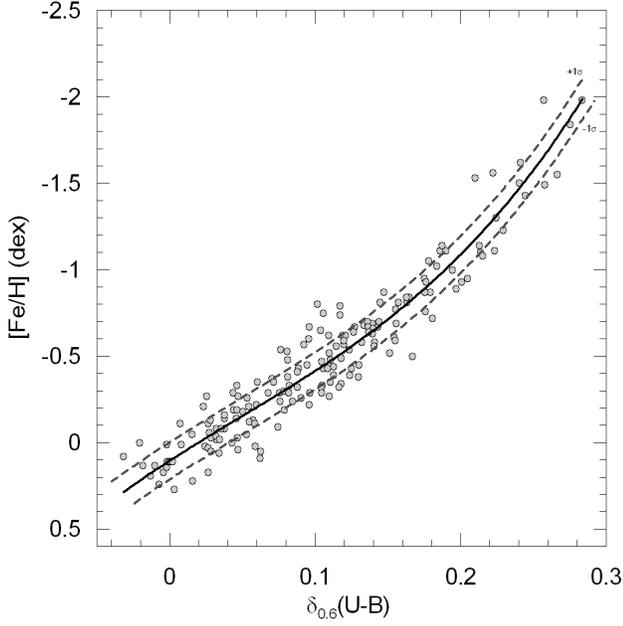}
\caption[] {Metallicity calibration. The solid line corresponds to the relation in Eq. (10). The dashed lines denote $\pm 1\sigma$ prediction levels.}
\end{center}
\end{figure}

\begin{figure}[h]
\begin{center}
\includegraphics[scale=0.4, angle=0]{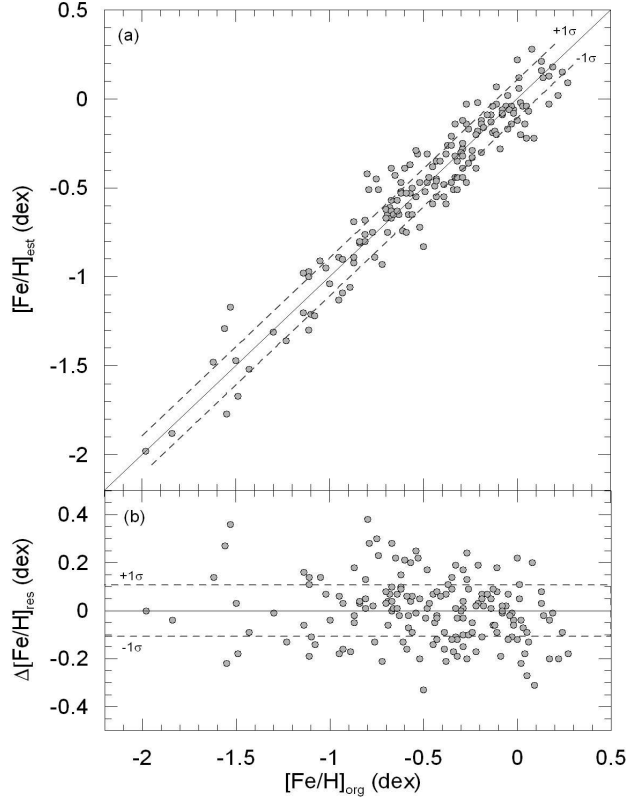}
\caption[] {Comparision of the original ($[Fe/H]_{org}$) and estimated ($[Fe/H]_{est}$) metallicities (a) and distribution of the metallicity residuals ($\Delta [Fe/H]_{res}$) respect to the original metallicities (b) for the sample stars. The dashed lines denote $\pm 1\sigma$ prediction levels.}
\end{center}
\end{figure}

Data used for the metallicity calibration are given in Table 2. We fitted the iron abundances (Table 1) in terms of normalized UV excesses to a third degree polynomial as follows and presented in Fig. 4:

\begin{eqnarray} 
[Fe/H]=0.105(0.010)-5.428(0.521)\times \delta_{0.6}\\ \nonumber
+7.340(5.895)\times \delta_{0.6}^2-50.081(16.359)\times \delta_{0.6}^3,\\ \nonumber
\end{eqnarray}
where $\delta_{0.6}$ represents $\delta_{0.6}(U-B)$. The correlation coefficient and standard deviation of the Eq. (10) are $R^2=0.949$ and $\sigma=0.134$ dex, respectively. 

We estimated the iron abundances of the sample stars, $[Fe/H]_{est}$, using the Eq. (10) and compared them with the original ones (Fig. 5). There is a linear relation between two sets of iron abundances. The mean and the standard deviation of the residuals (Table 2, Fig. 5) are small, $\langle\Delta[Fe/H]\rangle=0$ and $\sigma=0.134$ dex, confirming the accuracy of our procedure.

\subsection{Absolute Magnitude Calibration}
We preferred the widely used procedure mentioned in the introduction for the absolute magnitude calibration which is based on the offset from a standard main sequence. The absolute magnitude-colour diagram for the sample stars is given in Fig. 6 confronted to the main sequence of Hyades cluster. The iron abundances of the sample stars are also indicated with different symbols in this diagram, where a trend in the metallicity can be noticed in the direction of the absolute magnitude.         

\begin{figure}[t]
\begin{center}
\includegraphics[scale=0.4, angle=0]{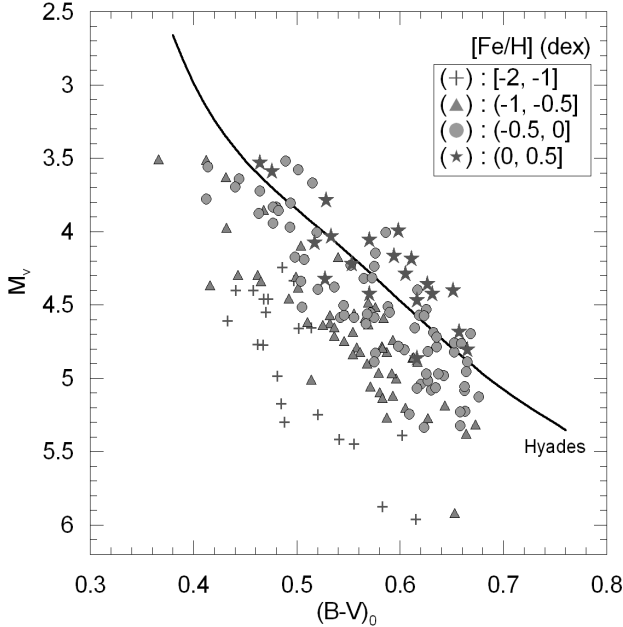}
\caption[] {Colour-absolute magnitude diagram for the sample stars. The curve denotes the intrinsic Hyades sequence. Symbols correspond to different metallicities as indicated in the panel.}
\end{center}
\end{figure}

\begin{figure}
\begin{center}
\includegraphics[scale=0.85, angle=0]{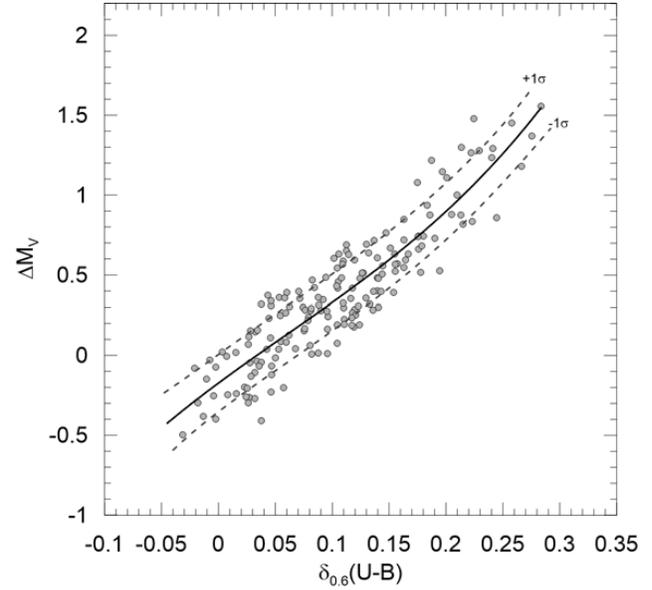}
\caption[] {Absolute magnitude offset versus normalized UV excess for the sample stars. The solid line corresponds to the relation in Eq. (11). The dashed lines denote $\pm 1\sigma$ prediction levels.}
\end{center}
\end{figure}

\begin{figure}
\begin{center}
\includegraphics[scale=0.4, angle=0]{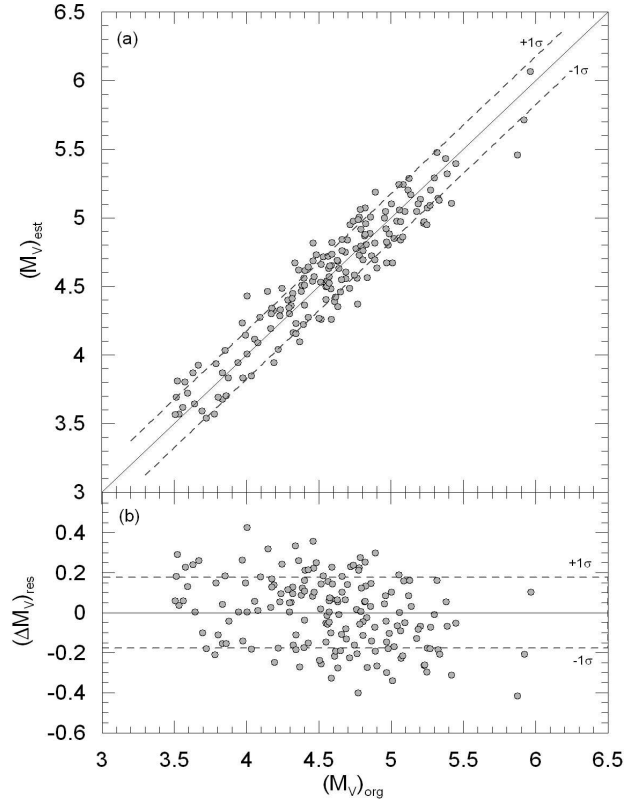}
\caption[] {Comparison of the original $(M_V)_{org}$ and estimated $(M_V)_{est}$ absolute magnitudes (a), and distribution of the absolute magnitude residuals $(\Delta M_V)_{res}$ respect to the original absolute magnitudes (b), for the sample stars. The dashed lines denote $\pm 1\sigma$ prediction levels.}
\end{center}
\end{figure}

We estimated the differences between the absolute magnitudes of the sample stars and the Hyades' stars of equal $(B-V)_0$ colour, $\Delta M_V=M_V-(M_V)_{Hya}$, and then plotted them against normalized UV excesses, $\delta_{0.6}(U-B)$, (Table 2 and Fig. 7) which yields the following third degree polynomial:

\begin{eqnarray}         
\Delta M_V=-0.174(0.036)+5.278(1.141)\times \delta_{0.6}\\ \nonumber
-5.292(1.007)\times \delta_{0.6}^2+28.477(5.237)\times \delta_{0.6}^3,\\ \nonumber        
\end{eqnarray}
The correlation coefficient and standard deviation of the Eq. (11) are $R^2=0.824$ and $\sigma=0.174$ mag, respectively. 

We estimated the absolute magnitudes of the sample stars, $(M_V)_{est}$, by replacing their normalized UV excesses into Eq. (11), and compared them with the original ones, $(M_V)_{org}$, to test the accuracy of the procedure. The comparison is shown in Fig. 8, and the residuals are listed in Table 2. As in the case of iron abundances, there is a linear relation between the two sets of absolute magnitudes, and the mean and the standard deviation of the residuals are small, $\langle\Delta M_V\rangle=0$ and $\sigma=0.174$ mag, confirming the accuracy of the procedure used in our study.          

\begin{figure*}[t]
\begin{center}
\includegraphics[scale=0.55, angle=0]{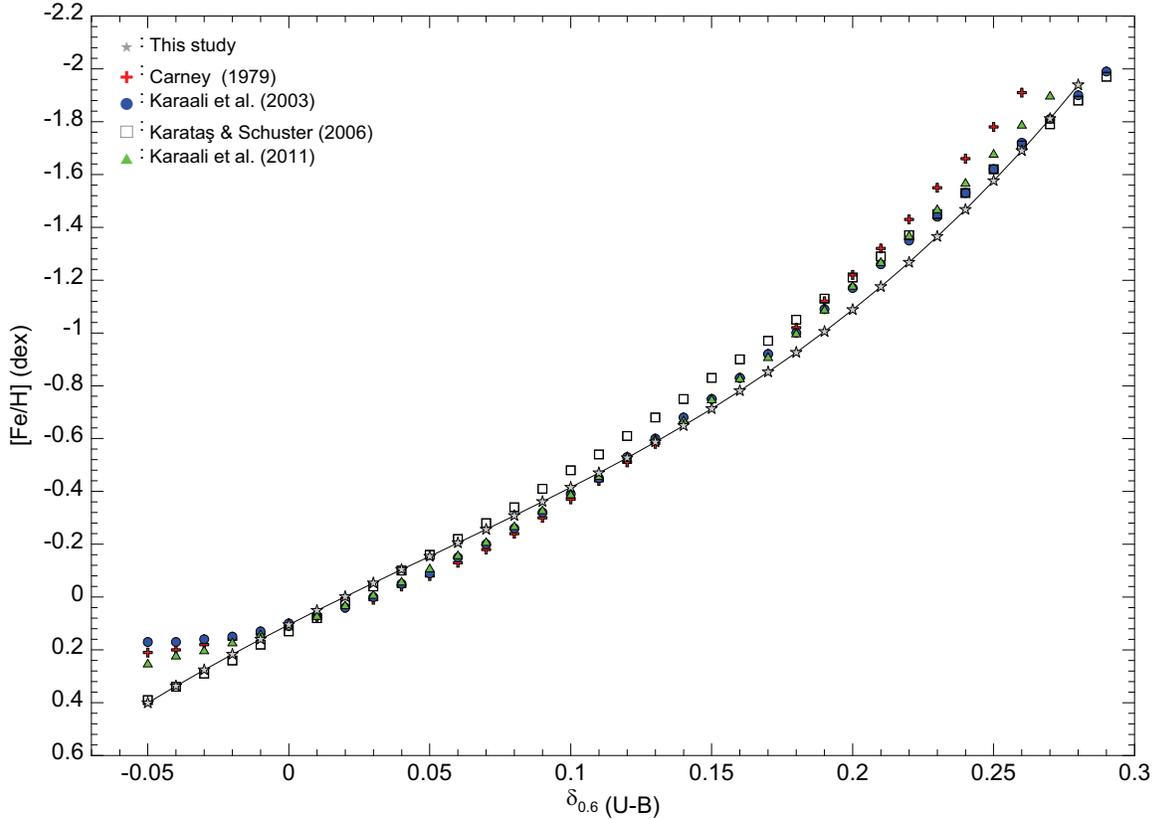}
\caption[] {Comparison of the metallicity calibrations for different studies as indicated in the panel.}
\end{center}
\end{figure*}

\begin{figure*}
\begin{center}
\includegraphics[scale=0.70, angle=0]{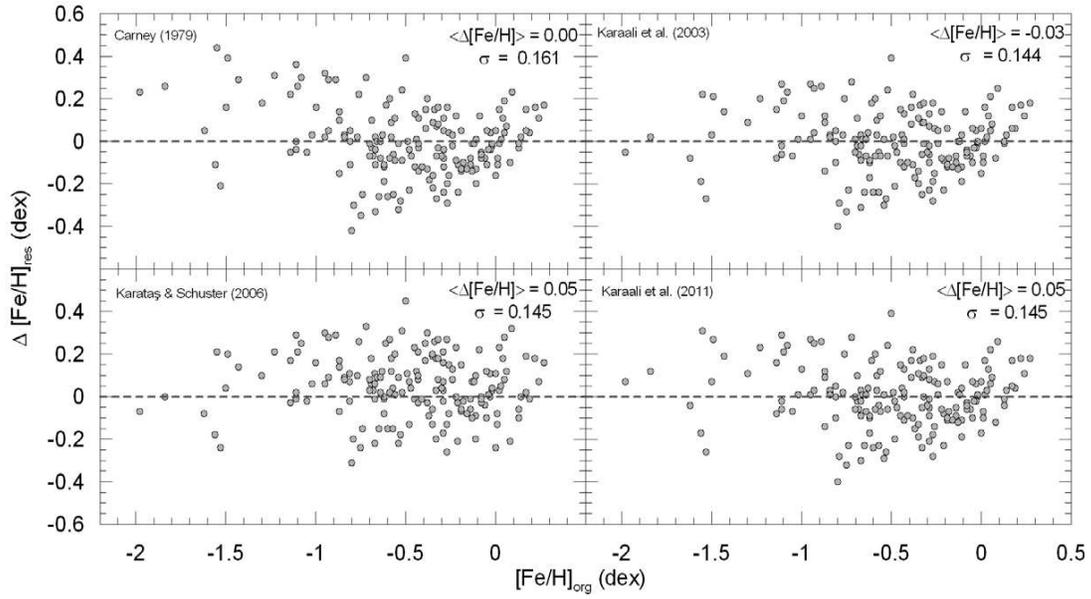}
\caption[] {Distribution of the metallicity residuals $(\Delta [Fe/H])_{res}$ respect to the original metallicities for four studies as indicated in the panels.}
\end{center}
\end{figure*}

\begin{figure*}
\begin{center}
\includegraphics[scale=0.45, angle=0]{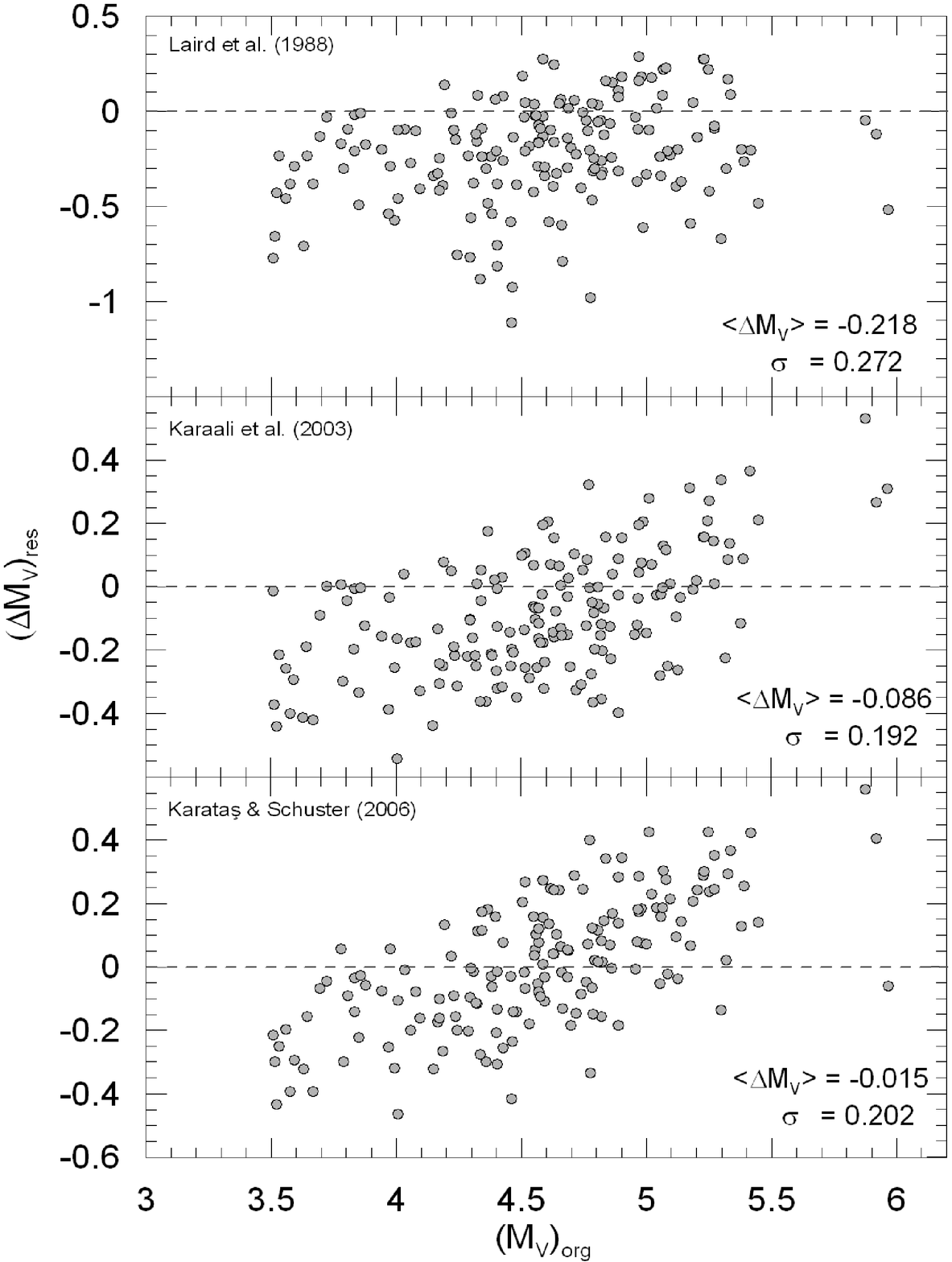}
\caption[] {Distribution of the absolute magnitude residuals $(\Delta M_V)_{res}$ respect to the original absolute magnitudes three studies as indicated in the panels.}
\end{center}
\end{figure*}

\subsection{Comparison with Other Analyses in the Literature}
The metallicity calibration is carried out with the same procedure descried in \citet{Carney79, Karaali03, Karaali11}, and \citet{Karatas06}. However, there are some differences between these calibrations due to different data. 

The metallicity calibrations in the literature and in this study are plotted in Fig. 9 with different symbols, i.e. ($+$): \citet{Carney79}, ($\bullet$): \citet{Karaali03}, ($\Box$): \citet{Karatas06}, ($\bigtriangleup$): \citet{Karaali11}, and ($\star$): this study. Our calibration provides richer metallicities, $0.0<\Delta [Fe/H]<0.1$ dex, for the metal-poor stars with $[Fe/H]<-0.6$ dex respect to the cited calibrations except the ones in \citet{Carney79} which give poorer metallicities compared to all calibrations mentioned in this study. Metal-rich part of the our calibration is compatible with the one in \citet{Karatas06} in the interval $0.2<[Fe/H]<0.4$ dex, and for the intermediate metallicities, $-0.6<[Fe/H]<0.2$ dex, our calibration curve occupies a central position with respect to the other ones.
  
We applied the four cited metallicity calibrations to the sample stars used in our study for a further comparison of our results with the ones in cited studies. We replaced their normalized UV excesses, $\delta_{0.6}(U-B)$, in the corresponding metallicity calibrations in \citet{Carney79}, \citet{Karaali03}, \citet{Karatas06}, and \citet{Karaali11}, and estimated the iron abundance, $[Fe/H]_{est}$, of each star. Then, we evaluated the residual metallicities, $\Delta[Fe/H]_{res}=[Fe/H]_{org}-[Fe/H]_{est}$, and plotted them in terms of $[Fe/H]_{org}$ in Fig. 10 for each study, where $[Fe/H]_{org}$ denotes the original iron abundances. The means and the corresponding standard deviations of the residuals are $\langle\Delta[Fe/H]\rangle=0$, -0.03, 0.05, and 0.05 dex; and $\sigma$=0.161, 0.144, 0.145, and 0.145 dex for the studies in the order given above. The mean of the residuals and their standard deviation, $\langle\Delta [Fe/H]\rangle=0$ and $\sigma=0.134$ dex, in our study are equal or smaller than the ones for the cited studies which indicate that the new data used in our study improved the metallicity calibration. 

A similar comparison is carried out for the absolute magnitudes. We applied the absolute magnitude calibrations, i.e. $\Delta M_V$ offsets versus normalized UV excesses $\delta_{0.6}(U-B)$, in \citet{Laird88}, \citet{Karaali03}, and \citet{Karatas06} to the sample stars in our study and estimated their absolute magnitudes $(M_V)_{est}$. Then, we evaluated the residuals $(\Delta M_V)_{res}=(M_V)_{est}-(M_V)_{org}$ and plotted them in terms of $(M_V)_{org}$ in Fig. 11, where $(M_V)_{org}$ denotes the original absolute magnitudes. The mean of the residuals and the corresponding standard deviations are $\Delta (M_V)_{res}=$ -0.218, -0.086, and -0.015 mag; and $\sigma$=0.272, 0.192, and 0.202 mag, for the cited studies in the order given above. The mean of the absolute magnitude residuals (0 mag) and their standard deviation (0.174 mag) in our study are (absolutely) much smaller than the corresponding ones in the cited studies which is a result of the improved data used in our study. 

\section{Summary and Discussion}
We used a sample of F-G dwarfs taken from different sources and calibrated their iron abundances and absolute magnitudes in terms of UV excesses. We used the re-reduced {\it Hipparcos} parallaxes \citep{vanLeeuwen07} for the first time and obtained accurate distances and absolute magnitudes. Our sample consists of 168 dwarf stars covering the colour, iron abundance and absolute magnitude intervals $0.30<(B-V)_0<0.68$ mag, $-2.0<[Fe/H]<0.4$ dex and $3.4<M_V<6.0$ mag, respectively. The means and standard deviations of the metallicity and absolute magnitude residuals are small, i.e. $\langle\Delta[Fe/H]_{res}\rangle=0$ and $\sigma=0.134$ dex, and $\langle\Delta (M_V)_{res}\rangle=0$ and $\sigma=0.174$ mag, respectively, which indicate accurate metallicity and photometric parallax estimations. 

$UBV$ is the oldest standard photometry in astronomy. Hence, many calibrations are carried out with this system. Although many other photometric systems such as Sloan Digital Sky Survey \citep[SDSS;][]{York00}, Two Micron All Sky Survey \citep[2MASS;][]{Skrutskie06} and Wide field Infrared Survey Explorer \citep[WISE;][]{Wright10} are defined on different spectral bands for different purposes, the $UBV$ is still widely used due to its advantage, i.e. apparently bright stars for which reliable trigonometric parallaxes and hence distances can be provided in this system. This is the main reason that we preferred the $UBV$ photometry in this study. However, we plan to extend this calibrations to other photometric systems which are widely used for other purposes. Our favorite photometric system is the SDSS which provide useful investigations in the Galactic structure. We can use the transformation equations in the literature and transform our calibrations to the SDSS data. Then, we can apply them to a larger sample of stars observed with SDSS photometry. This will be the subject of a second study.

\section{Acknowledgments}
Authors are grateful to the anonymous referee whose useful comments 
and improvements for the manuscript. This work has been supported in 
part by the Scientific and Technological Research Council 
(T\"UB\.ITAK) 114F347. Part of this work was supported by the Research 
Fund of the University of Istanbul, Project Number: 52265. This 
research has made use of the SIMBAD, and NASA\rq s Astrophysics Data 
System Bibliographic Services.

\begin{table*}
\setlength{\tabcolsep}{2pt} 
{\scriptsize
\caption{Data for the sample stars used in our study. The columns give: ID, Equatorial coordinates, Hipparcos number, $V$ apparent magnitude, $U-B$ and $B-V$ colours, colour excess reduced to the distance of the star ($E_d(B-V)$), parallax and its error ($\pi$), effective temperature ($T_{eff}$), surface gravity ($\log~g$), iron abundance ($[Fe/H]$), and reference.} 
\begin{center}
\begin{tabular}{ccccccccccccl}
\hline
        ID & $\alpha_{2000}$ &$\delta_{2000}$ &        Hip &        $V$ &      $U-B$ &      $B-V$ & $E_d(B-V)$ &      $\pi$ &  $T_{eff}$ &   $\log g$ &   $[Fe/H]$ &~~~~~~~~~Authors \\
         & (hh:mm:ss.ss)&(dd:mm:ss.s)&  &    (mag)   &      (mag) &      (mag) & (mag) &      (mas) &  (K) &   (cgs) &   (dex) &    \\
\hline
         1 & 00 05 54.70 &  +18 14 06.0 &        493 &      7.450 &      0.040 &      0.560 &      0.005 & 26.93$\pm$0.56 &       5826 &       4.36 &      -0.29 & 2004AJ....128.1177V \\
         2 & 00 11 15.86 &$-$15 28 04.7 &        910 &      4.895 &     -0.014 &      0.492 &      0.003 & 53.34$\pm$0.64 &       6289 &       4.17 &      -0.33 & 2014A\&A...562A..71B \\
         3 & 00 20 04.26 &$-$64 52 29.3 &       1599 &      4.228 &      0.018 &      0.572 &      0.001 & 116.46$\pm$0.60 &       5932 &       4.33 &      -0.19 & 2014A\&A...562A..71B \\
         4 & 00 25 01.42 &$-$30 41 51.7 &       1976 &      7.560 &      0.155 &      0.615 &      0.004 & 21.27$\pm$0.70 &       5982 &       4.32 &       0.19 & 2014A\&A...562A..71B \\
         5 & 00 34 27.83 &$-$52 22 23.1 &       2711 &      5.569 &     -0.016 &      0.466 &      0.002 & 39.24$\pm$0.34 &       6499 &       4.22 &       0.06 & 2014A\&A...562A..71B \\
         6 & 00 40 32.24 &$-$29 52 05.8 &       3182 &      8.720 &      0.090 &      0.635 &      0.008 & 16.76$\pm$0.98 &       5643 &       4.32 &      -0.29 & 2014A\&A...562A..71B \\
         7 & 00 44 26.65 &$-$26 30 56.4 &       3479 &      7.783 &      0.156 &      0.665 &      0.002 & 30.89$\pm$0.75 &       5563 &       4.40 &      -0.26 & 2014A\&A...562A..71B \\
         8 & 00 44 39.27 &$-$65 38 58.3 &       3497 &      6.544 &      0.107 &      0.654 &      0.002 & 45.34$\pm$0.32 &       5638 &       4.41 &      -0.32 & 2014A\&A...562A..71B \\
         9 & 00 47 30.75 &$-$36 56 24.7 &       3704 &      7.831 &     -0.041 &      0.540 &      0.004 & 20.51$\pm$0.78 &       6078 &       4.40 &      -0.35 & 2014A\&A...562A..71B \\
        10 & 00 50 07.59 &$-$10 38 39.6 &       3909 &      5.187 &     -0.003 &      0.510 &      0.003 & 63.48$\pm$0.35 &       6352 &       4.45 &      -0.03 & 2014A\&A...562A..71B \\
        11 & 00 58 11.69 &  +80 06 49.3 &       4544 &      9.780 &     -0.100 &      0.540 &      0.048 & 9.23$\pm$1.03 &       5832 &       4.51 &      -0.87 & 2006MNRAS.367.1329R \\
        12 & 01 02 49.72 &$-$37 18 58.2 &       4892 &      8.516 &      0.008 &      0.580 &      0.005 & 16.48$\pm$1.04 &       5994 &       4.42 &      -0.29 & 2014A\&A...562A..71B \\
        13 & 01 06 05.15 &  +01 42 23.1 &       5163 &      9.448 &      0.000 &      0.594 &      0.010 & 10.84$\pm$1.29 &       5547 &       4.63 &      -0.74 & 2006MNRAS.367.1329R \\
        14 & 01 07 48.66 &$-$08 14 01.3 &       5301 &      8.440 &      0.170 &      0.650 &      0.018 & 18.23$\pm$0.76 &       5686 &       4.39 &      -0.11 & 2014A\&A...562A..71B \\
        15 & 01 15 11.12 &$-$45 31 54.0 &       5862 &      4.957 &      0.098 &      0.571 &      0.001 & 66.16$\pm$0.24 &       6145 &       4.24 &       0.17 & 2014A\&A...562A..71B \\
        16 & 01 18 59.99 &$-$08 56 22.2 &       6159 &      8.898 &     -0.019 &      0.595 &      0.012 & 15.35$\pm$1.17 &       5653 &       4.56 &      -0.67 & 2006MNRAS.367.1329R \\
        17 & 01 32 57.60 &  +23 41 44.0 &       7217 &      9.040 &      0.062 &      0.624 &      0.026 & 14.60$\pm$1.14 &       5550 &       4.20 &      -0.48 & 2004AJ....128.1177V \\
        18 & 01 36 05.80 &$-$61 05 03.0 &       7459 &     10.100 &     -0.160 &      0.530 &      0.010 & 10.87$\pm$1.33 &       5759 &       4.31 &      -1.23 & 2010A\&A...511L..10N \\
        19 & 01 42 29.32 &$-$53 44 27.0 &       7978 &      5.540 &     -0.004 &      0.530 &      0.003 & 57.36$\pm$0.25 &       6219 &       4.41 &       0.05 & 2014A\&A...562A..71B \\
        20 & 01 53 57.68 &  +10 36 50.5 &       8859 &      6.780 &     -0.050 &      0.460 &      0.016 & 24.12$\pm$0.64 &       6420 &       4.21 &      -0.30 & 2014A\&A...562A..71B \\
        21 & 01 56 59.98 &$-$51 45 58.5 &       9085 &      6.097 &     -0.060 &      0.480 &      0.003 & 37.22$\pm$0.38 &       6334 &       4.29 &      -0.26 & 2014A\&A...562A..71B \\
        22 & 02 14 40.30 &$-$01 12 05.1 &      10449 &      9.081 &     -0.071 &      0.579 &      0.008 & 15.87$\pm$1.23 &       5566 &       4.64 &      -0.87 & 2006MNRAS.367.1329R \\
        23 & 02 17 07.14 &  +21 34 00.5 &      10652 &      9.042 &      0.012 &      0.620 &      0.028 & 15.93$\pm$1.19 &       5499 &       4.58 &      -0.67 & 2006MNRAS.367.1329R \\
        24 & 02 38 21.50 &  +02 26 44.4 &      12294 &     10.528 &     -0.158 &      0.462 &      0.019 & 5.82$\pm$2.18 &       6069 &       4.58 &      -0.95 & 2006MNRAS.367.1329R \\
        25 & 02 38 27.86 &  +30 48 59.9 &      12306 &      7.359 &     -0.003 &      0.586 &      0.017 & 29.17$\pm$0.81 &       5832 &       4.44 &      -0.52 & 2014A\&A...562A..71B \\
        26 & 02 40 12.42 &$-$09 27 10.4 &      12444 &      5.775 &     -0.011 &      0.520 &      0.003 & 45.96$\pm$0.41 &       6383 &       4.41 &       0.09 & 2014A\&A...562A..71B \\
        27 & 02 42 33.47 &$-$50 48 01.1 &      12653 &      5.403 &      0.075 &      0.556 &      0.003 & 58.25$\pm$0.22 &       6375 &       4.73 &       0.27 & 2014A\&A...562A..71B \\
        28 & 02 44 12.00 &  +49 13 42.0 &      12777 &      4.100 &     -0.007 &      0.485 &      0.003 & 89.87$\pm$0.22 &       5309 &       4.30 &      -0.02 & 2004AJ....128.1177V \\
        29 & 02 45 41.01 &$-$38 09 31.4 &      12889 &      7.637 &      0.046 &      0.580 &      0.004 & 20.16$\pm$0.69 &       5999 &       4.23 &      -0.13 & 2014A\&A...562A..71B \\
        30 & 02 51 58.36 &  +11 22 11.9 &      13366 &      8.383 &     -0.061 &      0.547 &      0.048 & 16.39$\pm$1.07 &       5856 &       4.26 &      -0.69 & 2014A\&A...562A..71B \\
        31 & 03 03 38.96 &$-$05 39 58.7 &      14241 &      8.085 &      0.117 &      0.670 &      0.012 & 28.54$\pm$0.97 &       5430 &       4.33 &      -0.45 & 2014A\&A...562A..71B \\
        32 & 03 15 06.39 &$-$45 39 53.4 &      15131 &      6.760 &     -0.020 &      0.580 &      0.004 & 41.34$\pm$0.40 &       5985 &       4.54 &      -0.43 & 2014A\&A...562A..71B \\
        33 & 03 15 22.52 &$-$01 10 43.1 &      15158 &      8.550 &      0.030 &      0.530 &      0.029 & 10.55$\pm$1.15 &       6191 &       4.18 &      -0.06 & 2014A\&A...562A..71B \\
        34 & 03 18 19.98 &  +18 10 17.8 &      15381 &      7.540 &      0.170 &      0.620 &      0.022 & 20.16$\pm$0.64 &       5882 &       4.24 &       0.08 & 2014A\&A...562A..71B \\
        35 & 03 28 21.08 &$-$06 31 51.3 &      16169 &      8.238 &      0.028 &      0.621 &      0.009 & 21.38$\pm$0.88 &       5650 &       4.34 &      -0.57 & 2014A\&A...562A..71B \\
        36 & 03 40 22.06 &$-$03 13 01.1 &      17147 &      6.689 &     -0.086 &      0.540 &      0.008 & 39.12$\pm$0.56 &       5970 &       4.52 &      -0.81 & 2014A\&A...562A..71B \\
        37 & 04 52 09.91 &$-$27 03 51.0 &      22632 &      9.134 &     -0.190 &      0.490 &      0.009 & 15.00$\pm$1.13 &       5909 &       4.22 &      -1.62 & 2014A\&A...562A..71B \\
        38 & 05 03 53.95 &$-$41 44 41.8 &      23555 &      6.302 &      0.040 &      0.530 &      0.002 & 31.50$\pm$0.31 &       6466 &       4.64 &       0.22 & 2014A\&A...562A..71B \\
        39 & 05 05 28.70 &  +40 15 26.0 &      23688 &      9.660 &     -0.160 &      0.420 &      0.004 & 8.79$\pm$1.20 &       6293 &       4.41 &      -0.89 & 2010A\&A...511L..10N \\
        40 & 05 09 56.96 &  +05 33 26.7 &      24030 &      9.720 &     -0.130 &      0.520 &      0.023 & 8.66$\pm$1.77 &       5738 &       4.64 &      -1.00 & 2006MNRAS.367.1329R \\
        41 & 05 31 13.78 &  +15 46 24.4 &      25860 &      8.635 &      0.118 &      0.669 &      0.042 & 20.07$\pm$1.15 &       5543 &       4.56 &      -0.35 & 2006MNRAS.367.1329R \\
        42 & 05 39 27.44 &  +03 57 02.7 &      26617 &     10.350 &      0.030 &      0.630 &      0.076 & 8.79$\pm$2.67 &       5429 &       4.67 &      -0.75 & 2006MNRAS.367.1329R \\
        43 & 05 44 27.79 &$-$22 26 54.2 &      27072 &      3.590 &     -0.007 &      0.481 &      0.001 & 112.02$\pm$0.18 &       6323 &       4.16 &      -0.02 & 2014A\&A...562A..71B \\
        44 & 05 59 55.79 &$-$37 03 24.2 &      28403 &      8.594 &      0.050 &      0.626 &      0.006 & 19.64$\pm$0.72 &       5668 &       4.31 &      -0.33 & 2014A\&A...562A..71B \\
        45 & 06 03 14.86 &  +19 21 38.7 &      28671 &      9.310 &     -0.043 &      0.618 &      0.016 & 16.81$\pm$2.04 &       5396 &       4.39 &      -1.11 & 2014A\&A...562A..71B \\
        46 & 06 12 00.60 &  +06 46 59.0 &      29432 &      6.850 &      0.125 &      0.633 &      0.008 & 42.55$\pm$0.55 &       5726 &       4.58 &      -0.16 & 2004AJ....128.1177V \\
        47 & 06 32 37.99 &$-$06 29 18.5 &      31188 &      8.622 &     -0.069 &      0.560 &      0.035 & 16.79$\pm$0.91 &       5789 &       4.54 &      -0.59 & 2006MNRAS.367.1329R \\
        48 & 06 58 38.54 &$-$00 28 49.7 &      33582 &      9.018 &     -0.017 &      0.579 &      0.010 & 12.56$\pm$1.16 &       5773 &       4.11 &      -0.62 & 2014A\&A...562A..71B \\
        49 & 07 03 30.46 &  +29 20 13.5 &      34017 &      5.930 &      0.060 &      0.591 &      0.003 & 52.27$\pm$0.41 &       5920 &       4.42 &      -0.12 & 2014A\&A...562A..71B \\
        50 & 07 09 04.96 &  +15 25 17.7 &      34511 &      8.000 &      0.120 &      0.620 &      0.006 & 21.64$\pm$0.79 &       5789 &       4.44 &      -0.11 & 2014A\&A...562A..71B \\
        51 & 07 15 50.80 &$-$13 02 58.1 &      35139 &      7.755 &     -0.017 &      0.607 &      0.002 & 30.95$\pm$0.75 &       5771 &       4.47 &      -0.52 & 2014A\&A...562A..71B \\
        52 & 07 30 29.02 &  +18 57 40.6 &      36491 &      8.494 &     -0.117 &      0.518 &      0.004 & 20.20$\pm$1.29 &       5869 &       4.31 &      -0.95 & 2014A\&A...562A..71B \\
        53 & 07 34 35.11 &  +16 54 04.0 &      36849 &      8.950 &     -0.110 &      0.510 &      0.009 & 12.37$\pm$1.23 &       6012 &       4.16 &      -0.84 & 2014A\&A...562A..71B \\
        54 & 07 40 54.38 &$-$26 21 48.6 &      37419 &      8.683 &      0.070 &      0.625 &      0.009 & 19.16$\pm$0.88 &       5715 &       4.38 &      -0.33 & 2014A\&A...562A..71B \\
        55 & 07 53 33.12 &  +30 36 18.3 &      38541 &      8.300 &     -0.125 &      0.619 &      0.004 & 34.30$\pm$0.90 &       5316 &       4.60 &      -1.84 & 2014A\&A...562A..71B \\
        56 & 07 56 10.20 &  +50 32 27.3 &      38769 &      8.830 &     -0.060 &      0.520 &      0.016 & 11.56$\pm$0.98 &       5726 &       4.15 &      -0.79 & 2006MNRAS.367.1329R \\
        57 & 08 11 38.64 &  +32 27 25.7 &      40118 &      6.827 &      0.127 &      0.679 &      0.003 & 45.9$\pm$0.55 &       5484 &       4.58 &      -0.47 & 2006MNRAS.367.1329R \\
        58 & 08 19 22.60 &  +54 05 10.0 &      40778 &      9.741 &     -0.212 &      0.480 &      0.013 & 10.35$\pm$1.56 &       6006 &       4.23 &      -1.55 & 2010A\&A...511L..10N \\
        59 & 08 27 36.80 &  +45 39 11.0 &      41484 &      6.320 &      0.120 &      0.625 &      0.002 & 44.94$\pm$0.46 &       5703 &       4.46 &      -0.08 & 2004AJ....128.1177V \\
        60 & 08 38 08.51 &  +26 02 56.3 &      42356 &      7.579 &      0.135 &      0.632 &      0.006 & 22.91$\pm$0.59 &       5962 &       4.42 &       0.17 & 2014A\&A...562A..71B \\
        61 & 08 52 44.51 &  +22 33 30.2 &      43595 &     10.830 &     -0.040 &      0.600 &      0.013 & 7.87$\pm$2.47 &       5506 &       4.67 &      -0.84 & 2006MNRAS.367.1329R \\
        62 & 08 54 17.90 &$-$05 26 04.0 &      43726 &      6.010 &      0.212 &      0.666 &      0.001 & 57.52$\pm$0.39 &       5763 &       4.37 &       0.01 & 2004AJ....128.1177V \\
        63 & 08 59 06.00 &$-$00 37 26.0 &      44116 &      8.480 &     -0.103 &      0.440 &      0.008 & 12.70$\pm$1.15 &       6275 &       4.10 &      -0.58 & 2004AJ....128.1177V \\
        64 & 09 06 38.83 &$-$43 29 31.1 &      44713 &      7.307 &      0.202 &      0.666 &      0.015 & 26.83$\pm$0.51 &       5823 &       4.34 &       0.14 & 2014A\&A...562A..71B \\
        65 & 09 07 56.60 &$-$50 28 57.0 &      44811 &      7.720 &     -0.040 &      0.564 &      0.032 & 24.53$\pm$0.56 &       5824 &       4.45 &      -0.64 & 2004AJ....128.1177V \\
        66 & 09 35 16.70 &$-$49 07 48.9 &      47048 &      8.550 &     -0.095 &      0.398 &      0.032 & 10.26$\pm$0.95 &       6630 &       4.12 &      -0.50 & 2014A\&A...562A..71B \\
        67 & 09 36 49.50 &  +57 54 41.0 &      47174 &      9.960 &      0.030 &      0.630 &      0.007 & 11.99$\pm$1.71 &       5603 &       4.33 &      -0.45 & 2010A\&A...511L..10N \\
        68 & 09 49 42.82 &  +65 18 15.0 &      48209 &      9.696 &      0.100 &      0.674 &      0.031 & 13.10$\pm$1.20 &       5415 &       4.64 &      -0.65 & 2006MNRAS.367.1329R \\
        69 & 10 03 37.38 &$-$29 02 36.0 &      49285 &      8.110 &      0.200 &      0.680 &      0.012 & 21.13$\pm$1.06 &       5520 &       4.29 &      -0.21 & 2014A\&A...562A..71B \\
        70 & 10 07 33.80 &$-$06 26 21.0 &      49615 &      7.720 &     -0.060 &      0.510 &      0.006 & 21.25$\pm$0.62 &       6019 &       4.29 &      -0.43 & 2004AJ....128.1177V \\
        71 & 10 09 49.58 &$-$36 45 14.9 &      49793 &      8.082 &     -0.020 &      0.596 &      0.009 & 22.58$\pm$0.69 &       5764 &       4.33 &      -0.63 & 2014A\&A...562A..71B \\
        72 & 10 11 48.07 & +23 45 18.7 &      49942 &      8.419 &      0.115 &      0.632 &      0.007 & 16.86$\pm$0.77 &       5676 &       4.36 &      -0.27 & 2014A\&A...562A..71B \\
        73 & 10 22 46.93 &$-$45 28 14.2 &      50834 &      9.320 &     -0.070 &      0.438 &      0.026 & 8.09$\pm$1.12 &       6552 &       4.23 &      -0.27 & 2014A\&A...562A..71B \\
        74 & 10 24 35.68 &$-$05 31 10.8 &      50965 &      9.800 &      0.000 &      0.570 &      0.015 & 9.14$\pm$1.37 &       5715 &       4.56 &      -0.57 & 2006MNRAS.367.1329R \\
        75 & 10 46 14.24 &$-$29 20 25.5 &      52673 &      9.570 &      0.030 &      0.640 &      0.013 & 14.07$\pm$1.29 &       5541 &       4.62 &      -0.66 & 2006MNRAS.367.1329R \\
        76 & 10 47 23.16 &  +28 23 55.9 &      52771 &     10.240 &     -0.223 &      0.500 &      0.012 & 10.45$\pm$1.42 &       5354 &       4.77 &      -1.98 & 2006MNRAS.367.1329R \\
        77 & 10 57 09.60 &  +21 48 17.0 &      53537 &      7.940 &      0.120 &      0.620 &      0.004 & 20.32$\pm$0.66 &       6019 &       4.43 &       0.05 & 2004AJ....128.1177V \\
        78 & 10 59 28.00 &  +40 25 49.0 &      53721 &      5.030 &      0.124 &      0.606 &      0.001 & 71.11$\pm$0.25 &       5882 &       4.34 &       0.01 & 2004AJ....128.1177V \\
        79 & 11 08 40.07 &$-$44 15 33.7 &      54469 &      9.814 &     -0.020 &      0.563 &      0.021 & 9.27$\pm$1.29 &       6098 &       4.34 &      -0.43 & 2014A\&A...562A..71B \\
        80 & 11 11 00.74 &$-$65 25 37.8 &      54641 &      8.160 &     -0.158 &      0.483 &      0.025 & 18.36$\pm$0.58 &       6168 &       4.29 &      -1.14 & 2014A\&A...562A..71B \\
        81 & 11 14 49.93 &$-$23 38 47.9 &      54924 &      9.068 &      0.004 &      0.584 &      0.016 & 15.02$\pm$1.09 &       5682 &       4.34 &      -0.80 & 2014A\&A...562A..71B \\
        82 & 11 23 16.23 &  +19 53 37.7 &      55592 &      9.973 &     -0.140 &      0.480 &      0.010 & 8.35$\pm$1.58 &       5995 &       4.14 &      -1.02 & 2014A\&A...562A..71B \\
        83 & 11 25 31.80 &  +42 37 58.0 &      55761 &      7.860 &     -0.040 &      0.540 &      0.004 & 22.95$\pm$0.64 &       5741 &       4.35 &      -0.62 & 2004AJ....128.1177V \\
        84 & 11 37 08.12 &$-$39 28 12.0 &      56664 &      8.717 &     -0.090 &      0.438 &      0.026 & 9.45$\pm$1.07 &       6364 &       4.07 &      -0.68 & 2014A\&A...562A..71B \\
        85 & 11 41 22.48 &$-$26 40 01.9 &      57017 &      7.530 &     -0.060 &      0.470 &      0.007 & 18.78$\pm$0.61 &       6380 &       4.35 &      -0.43 & 2014A\&A...562A..71B \\
\hline
\end{tabular}
\end{center}
}
\end{table*}

\begin{table*}
\setcounter{table}{1}
\setlength{\tabcolsep}{2pt} 
{\scriptsize
\begin{center}
\begin{tabular}{ccccccccccccl}
\hline
        ID & $\alpha_{2000}$ &$\delta_{2000}$ &        Hip &        $V$ &      $U-B$ &      $B-V$ & $E_d(B-V)$ &      $\pi$ &  $T_{eff}$ &   $\log g$ &   $[Fe/H]$ &~~~~~~~~~Authors \\
         & (hh:mm:ss.ss)&(dd:mm:ss.s)&  &    (mag)   &      (mag) &      (mag) & (mag) &      (mas) &  (K) &   (cgs) &   (dex) &    \\
\hline
        86 & 11 44 35.70 &  +25 32 12.0 &      57265 &     10.370 &     -0.131 &      0.479 &      0.017 & 6.25$\pm$1.71 &       5928 &       4.23 &      -0.93 & 2010A\&A...511L..10N \\
        87 & 11 46 31.10 &$-$40 30 01.0 &      57443 &      4.890 &      0.100 &      0.665 &      0.003 & 108.45$\pm$0.22 &  5524     &       4.29 &      -0.35 & 2004AJ....128.1177V \\
        88 & 11 46 35.15 &  +50 52 54.7 &      57450 &      9.918 &     -0.150 &      0.560 &      0.005 & 12.85$\pm$1.33 &       5315 &       4.74 &      -1.50 & 2006MNRAS.367.1329R \\
        89 & 12 04 05.56 &  +03 20 26.7 &      58843 &      9.219 &     -0.053 &      0.587 &      0.008 & 14.24$\pm$1.33 &       5726 &       4.41 &      -0.84 & 2014A\&A...562A..71B \\
        90 & 12 05 13.41 &$-$28 43 02.0 &      58950 &      7.782 &      0.140 &      0.650 &      0.008 & 27.85$\pm$0.70 &       5686 &       4.45 &      -0.19 & 2014A\&A...562A..71B \\
        91 & 12 10 57.93 &$-$46 19 19.1 &      59380 &      7.524 &     -0.054 &      0.561 &      0.007 & 27.31$\pm$0.65 &       5901 &       4.38 &      -0.58 & 2014A\&A...562A..71B \\
        92 & 12 30 50.10 &  +53 04 36.0 &      61053 &      6.200 &      0.015 &      0.547 &      0.002 & 45.92$\pm$0.35 &       6060 &       4.35 &      -0.11 & 2004AJ....128.1177V \\
        93 & 12 37 39.13 &  +38 41 03.7 &      61619 &      7.350 &     -0.020 &      0.480 &      0.004 & 17.81$\pm$0.46 &       6429 &       4.26 &       0.04 & 2014A\&A...562A..71B \\
        94 & 12 43 43.22 &$-$44 40 31.6 &      62108 &      9.909 &     -0.160 &      0.463 &      0.022 & 8.17$\pm$1.43 &       6293 &       4.37 &      -1.53 & 2014A\&A...562A..71B \\
        95 & 12 44 59.40 &  +39 16 44.0 &      62207 &      5.950 &     -0.044 &      0.548 &      0.002 & 57.55$\pm$0.32 &       5795 &       4.15 &      -0.59 & 2004AJ....128.1177V \\
        96 & 12 52 11.64 &$-$56 34 28.0 &      62809 &      8.477 &     -0.020 &      0.613 &      0.017 & 20.67$\pm$1.04 &       5658 &       4.38 &      -0.77 & 2014A\&A...562A..71B \\
        97 & 13 11 21.40 &  +09 37 33.5 &      64345 &      8.743 &      0.053 &      0.621 &      0.008 & 16.88$\pm$1.04 &       5598 &       4.37 &      -0.60 & 2014A\&A...562A..71B \\
        98 & 13 11 52.40 &  +27 52 41.0 &      64394 &      4.230 &      0.068 &      0.571 &      0.001 & 109.54$\pm$0.17 &      6029 &       4.38 &       0.03 & 2004AJ....128.1177V \\
        99 & 13 15 36.97 &  +09 00 57.7 &      64698 &      8.450 &      0.170 &      0.660 &      0.008 & 18.48$\pm$1.10 &       5624 &       4.35 &      -0.13 & 2014A\&A...562A..71B \\
       100 & 13 16 11.25 &  +35 53 09.1 &      64747 &      8.280 &      0.120 &      0.640 &      0.003 & 21.87$\pm$0.8 &       5718 &       4.47 &      -0.21 & 2014A\&A...562A..71B \\
       101 & 13 24 30.60 &  +20 27 22.0 &      65418 &     12.190 &     -0.174 &      0.450 &      0.017 & 3.12$\pm$3.30 &       6043 &       4.18 &      -1.56 & 2010A\&A...511L..10N \\
       102 & 13 34 32.65 &$-$38 54 26.0 &      66238 &      7.294 &      0.121 &      0.670 &      0.005 & 33.27$\pm$0.61 &       5660 &       4.42 &      -0.22 & 2014A\&A...562A..71B \\
       103 & 13 51 40.40 &$-$57 26 08.0 &      67655 &      7.970 &      0.005 &      0.662 &      0.009 & 39.42$\pm$0.97 &       5396 &       4.38 &      -0.93 & 2004AJ....128.1177V \\
       104 & 13 53 58.12 &$-$46 32 19.5 &      67863 &      9.038 &     -0.047 &      0.604 &      0.011 & 16.70$\pm$1.24 &       5692 &       4.35 &      -0.76 & 2014A\&A...562A..71B \\
       105 & 13 55 50.00 &  +14 03 23.0 &      68030 &      6.160 &     -0.085 &      0.501 &      0.003 & 40.22$\pm$0.37 &       6081 &       4.40 &      -0.38 & 2004AJ....128.1177V \\
       106 & 14 23 15.30 &  +01 14 30.0 &      70319 &      6.250 &      0.083 &      0.637 &      0.003 & 58.17$\pm$0.53 &       5597 &       4.44 &      -0.41 & 2004AJ....128.1177V \\
       107 & 14 27 24.91 &$-$18 24 40.4 &      70681 &      9.302 &     -0.103 &      0.597 &      0.014 & 21.04$\pm$1.12 &       5484 &       4.48 &      -1.30 & 2014A\&A...562A..71B \\
       108 & 14 29 03.26 &$-$46 44 28.0 &      70829 &      8.937 &      0.092 &      0.684 &      0.011 & 19.18$\pm$1.21 &       5471 &       4.47 &      -0.67 & 2014A\&A...562A..71B \\
       109 & 14 32 04.50 &  +18 50 10.0 &      71076 &      7.810 &     -0.080 &      0.500 &      0.007 & 17.24$\pm$0.75 &       6038 &       4.22 &      -0.39 & 2004AJ....128.1177V \\
       110 & 14 40 28.26 &$-$57 01 46.4 &      71735 &      7.385 &      0.150 &      0.670 &      0.012 & 37.71$\pm$0.72 &       5513 &       4.43 &      -0.35 & 2014A\&A...562A..71B \\
       111 & 14 48 18.70 &  +58 54 36.1 &      72407 &      9.760 &      0.073 &      0.620 &      0.003 & 10.65$\pm$0.73 &       5567 &       4.56 &      -0.54 & 2006MNRAS.367.1329R \\
       112 & 14 51 31.72 &$-$60 55 50.8 &      72673 &      7.176 &     -0.103 &      0.450 &      0.019 & 20.06$\pm$0.71 &       6374 &       4.29 &      -0.56 & 2014A\&A...562A..71B \\
       113 & 15 08 12.57 &$-$07 54 47.5 &      74067 &      8.000 &     -0.050 &      0.590 &      0.010 & 26.62$\pm$0.86 &       5695 &       4.38 &      -0.81 & 2014A\&A...562A..71B \\
       114 & 15 21 48.15 &$-$48 19 03.5 &      75181 &      5.651 &      0.069 &      0.639 &      0.004 & 67.51$\pm$0.39 &       5698 &       4.46 &      -0.32 & 2014A\&A...562A..71B \\
       115 & 16 03 13.30 &  +42 14 46.6 &      78640 &      9.854 &     -0.198 &      0.474 &      0.006 & 8.41$\pm$1.02 &       6069 &       4.30 &      -1.43 & 2014A\&A...562A..71B \\
       116 & 16 39 04.14 &$-$58 15 29.5 &      81520 &      7.021 &      0.029 &      0.615 &      0.006 & 44.54$\pm$0.54 &       5746 &       4.62 &      -0.44 & 2014A\&A...562A..71B \\
       117 & 16 41 08.21 &$-$02 51 26.2 &      81681 &      7.245 &      0.070 &      0.631 &      0.027 & 33.82$\pm$0.58 &       5565 &       4.52 &      -0.38 & 2006MNRAS.367.1329R \\
       118 & 17 00 31.65 &$-$57 17 49.6 &      83229 &      7.006 &     -0.020 &      0.574 &      0.006 & 32.77$\pm$0.63 &       5865 &       4.41 &      -0.49 & 2014A\&A...562A..71B \\
       119 & 17 03 49.15 &  +17 11 21.1 &      83489 &      9.128 &      0.100 &      0.651 &      0.016 & 13.43$\pm$1.41 &       5722 &       4.28 &      -0.29 & 2014A\&A...562A..71B \\
       120 & 17 22 12.65 &$-$75 20 53.3 &      84988 &      6.996 &     -0.012 &      0.599 &      0.006 & 35.67$\pm$0.50 &       5686 &       4.25 &      -0.70 & 2014A\&A...562A..71B \\
       121 & 17 22 27.65 &  +24 52 46.0 &      85007 &      6.870 &     -0.064 &      0.511 &      0.006 & 34.12$\pm$0.57 &       6064 &       4.32 &      -0.39 & 2014A\&A...562A..71B \\
       122 & 17 22 51.29 &$-$02 23 17.4 &      85042 &      6.282 &      0.235 &      0.680 &      0.021 & 51.22$\pm$0.40 &       5648 &       4.46 &       0.00 & 2014A\&A...562A..71B \\
       123 & 17 34 43.06 &  +06 00 51.6 &      86013 &      8.390 &     -0.050 &      0.570 &      0.012 & 19.38$\pm$1.14 &       5760 &       4.30 &      -0.81 & 2014A\&A...562A..71B \\
       124 & 17 38 15.61 &  +18 33 25.5 &      86321 &      9.781 &     -0.104 &      0.484 &      0.019 & 8.38$\pm$1.29 &       5760 &       4.59 &      -0.87 & 2006MNRAS.367.1329R \\
       125 & 17 47 28.00 &$-$08 46 48.0 &      87062 &     10.593 &     -0.133 &      0.590 &      0.105 & 9.59$\pm$2.21 &       6027 &       4.32 &      -1.49 & 2010A\&A...511L..10N \\
       126 & 18 09 21.38 &  +29 57 06.2 &      88945 &      6.850 &      0.090 &      0.620 &      0.004 & 40.29$\pm$0.49 &       5869 &       4.62 &       0.02 & 2014A\&A...562A..71B \\
       127 & 18 48 16.40 &  +23 30 53.1 &      92270 &      6.120 &      0.020 &      0.501 &      0.007 & 34.78$\pm$0.41 &       6376 &       4.23 &      -0.05 & 2014A\&A...562A..71B \\
       128 & 18 51 25.20 &  +38 37 36.0 &      92532 &      7.150 &     -0.045 &      0.540 &      0.004 & 32.72$\pm$0.36 &       5825 &       4.30 &      -0.56 & 2004AJ....128.1177V \\
       129 & 18 54 23.20 &$-$04 36 18.6 &      92781 &      9.071 &     -0.050 &      0.584 &      0.023 & 14.59$\pm$1.29 &       5765 &       4.33 &      -0.69 & 2014A\&A...562A..71B \\
       130 & 18 58 51.00 &  +30 10 50.3 &      93185 &      6.790 &      0.000 &      0.580 &      0.005 & 41.94$\pm$0.47 &       5864 &       4.54 &      -0.29 & 2014A\&A...562A..71B \\
       131 & 19 09 39.24 &$-$21 28 10.8 &      94129 &      8.210 &      0.130 &      0.630 &      0.013 & 17.6$\pm$0.76 &       5630 &       4.35 &      -0.27 & 2006MNRAS.367.1329R \\
       132 & 19 15 33.23 &$-$24 10 45.7 &      94645 &      6.254 &      0.070 &      0.540 &      0.007 & 36.30$\pm$0.70 &       6365 &       4.56 &       0.24 & 2014A\&A...562A..71B \\
       133 & 19 32 40.33 &$-$28 01 11.3 &      96124 &      7.151 &      0.155 &      0.671 &      0.007 & 36.72$\pm$0.95 &       5577 &       4.47 &      -0.22 & 2014A\&A...562A..71B \\
       134 & 19 34 19.79 &  +51 14 11.8 &      96258 &      5.733 &     -0.009 &      0.468 &      0.004 & 39.82$\pm$0.2 &       6380 &       4.15 &      -0.03 & 2014A\&A...562A..71B \\
       135 & 19 58 58.54 &$-$46 05 17.0 &      98355 &      7.473 &     -0.086 &      0.477 &      0.009 & 19.11$\pm$0.84 &       6232 &       4.18 &      -0.62 & 2014A\&A...562A..71B \\
       136 & 20 07 36.91 &$-$41 01 09.6 &      99139 &      8.840 &      0.100 &      0.640 &      0.010 & 17.95$\pm$1.16 &       5612 &       4.32 &      -0.37 & 2014A\&A...562A..71B \\
       137 & 20 16 38.08 &$-$07 26 37.8 &      99938 &      8.389 &     -0.024 &      0.575 &      0.009 & 17.64$\pm$0.82 &       5732 &       4.48 &      -0.54 & 2006MNRAS.367.1329R \\
       138 & 20 17 31.30 &  +66 51 13.0 &     100017 &      5.910 &      0.058 &      0.585 &      0.018 & 56.92$\pm$0.24 &       5782 &       4.44 &      -0.19 & 2004AJ....128.1177V \\
       139 & 20 20 24.60 &  +06 01 53.0 &     100279 &     10.134 &     -0.048 &      0.614 &      0.031 & 10.46$\pm$1.60 &       5673 &       4.31 &      -0.72 & 2010A\&A...511L..10N \\
       140 & 20 23 35.85 &$-$21 22 14.2 &     100568 &      8.653 &     -0.133 &      0.549 &      0.008 & 22.78$\pm$1.00 &       5800 &       4.48 &      -1.10 & 2014A\&A...562A..71B \\
       141 & 20 26 11.92 &  +09 27 00.4 &     100792 &      8.344 &     -0.174 &      0.483 &      0.011 & 17.00$\pm$0.83 &       6002 &       4.31 &      -1.11 & 2014A\&A...562A..71B \\
       142 & 20 40 22.33 &$-$24 07 04.9 &     102018 &      7.212 &      0.143 &      0.602 &      0.008 & 24.89$\pm$0.57 &       5933 &       4.12 &       0.13 & 2014A\&A...562A..71B \\
       143 & 20 40 49.38 &$-$18 47 33.3 &     102046 &      8.232 &     -0.131 &      0.495 &      0.009 & 16.15$\pm$0.93 &       6009 &       4.13 &      -1.05 & 2014A\&A...562A..71B \\
       144 & 20 49 15.19 &$-$20 37 50.8 &     102762 &      8.126 &      0.080 &      0.593 &      0.018 & 17.11$\pm$0.88 &       5948 &       4.30 &      -0.04 & 2014A\&A...562A..71B \\
       145 & 20 49 37.80 &  +12 32 42.0 &     102805 &      6.010 &     -0.090 &      0.420 &      0.006 & 32.66$\pm$0.41 &       6337 &       4.31 &      -0.32 & 2004AJ....128.1177V \\
       146 & 20 57 40.07 &$-$44 07 45.7 &     103458 &      6.519 &     -0.042 &      0.586 &      0.004 & 45.17$\pm$0.46 &       5843 &       4.53 &      -0.61 & 2014A\&A...562A..71B \\
       147 & 20 58 08.52 &$-$48 12 13.5 &     103498 &      8.290 &     -0.130 &      0.520 &      0.006 & 18.95$\pm$0.76 &       5856 &       4.23 &      -1.11 & 2014A\&A...562A..71B \\
       148 & 21 03 06.10 &  +29 28 56.0 &     103897 &     10.190 &      0.030 &      0.610 &      0.034 & 7.70$\pm$1.63 &       5607 &       4.39 &      -0.67 & 2010A\&A...511L..10N \\
       149 & 21 11 59.03 &  +17 43 39.9 &     104659 &      7.371 &     -0.158 &      0.511 &      0.009 & 29.10$\pm$0.64 &       5973 &       4.35 &      -1.08 & 2014A\&A...562A..71B \\
       150 & 21 43 57.12 &  +27 23 24.0 &     107294 &     10.050 &     -0.140 &      0.480 &      0.018 & 9.03$\pm$1.68 &       5929 &       4.63 &      -1.14 & 2006MNRAS.367.1329R \\
       151 & 21 51 24.61 &$-$23 16 14.2 &     107877 &      6.872 &     -0.044 &      0.484 &      0.007 & 24.91$\pm$0.56 &       6355 &       4.20 &      -0.24 & 2014A\&A...562A..71B \\
       152 & 21 58 24.32 &$-$12 39 52.8 &     108468 &      7.210 &      0.115 &      0.625 &      0.007 & 29.93$\pm$0.74 &       5799 &       4.25 &      -0.08 & 2014A\&A...562A..71B \\
       153 & 22 01 36.52 &$-$53 05 36.9 &     108736 &      7.100 &      0.025 &      0.569 &      0.004 & 27.95$\pm$0.55 &       5990 &       4.47 &      -0.29 & 2014A\&A...562A..71B \\
       154 & 22 06 33.17 &  +01 51 25.7 &     109144 &      7.244 &      0.020 &      0.524 &      0.009 & 19.52$\pm$0.77 &       6272 &       4.21 &      -0.08 & 2014A\&A...562A..71B \\
       155 & 22 09 34.61 &$-$41 13 29.6 &     109381 &      7.837 &      0.210 &      0.660 &      0.003 & 23.53$\pm$0.76 &       5803 &       4.46 &       0.13 & 2014A\&A...562A..71B \\
       156 & 22 12 43.50 &$-$06 28 08.0 &     109646 &      7.440 &     -0.080 &      0.520 &      0.010 & 27.64$\pm$0.68 &       5910 &       4.25 &      -0.64 & 2004AJ....128.1177V \\
       157 & 22 17 15.14 &  +12 53 54.6 &     110035 &      7.041 &      0.076 &      0.600 &      0.010 & 32.22$\pm$0.52 &       5907 &       4.42 &      -0.14 & 2014A\&A...562A..71B \\
       158 & 22 20 55.80 &  +08 11 12.3 &     110341 &      6.178 &     -0.052 &      0.450 &      0.010 & 32.33$\pm$0.55 &       6577 &       4.35 &      -0.09 & 2014A\&A...562A..71B \\
       159 & 22 23 49.10 &  +24 23 33.0 &     110560 &     10.640 &      0.009 &      0.571 &      0.031 & 5.32$\pm$1.65 &       5791 &       4.14 &      -0.53 & 2010A\&A...511L..10N \\
       160 & 22 36 07.70 &$-$54 36 38.2 &     111565 &      7.580 &      0.100 &      0.665 &      0.003 & 31.82$\pm$0.63 &       5534 &       4.44 &      -0.48 & 2014A\&A...562A..71B \\
       161 & 22 50 45.94 &  +01 51 54.6 &     112811 &      9.338 &      0.082 &      0.682 &      0.018 & 16.56$\pm$1.22 &       5347 &       4.64 &      -0.70 & 2006MNRAS.367.1329R \\
       162 & 23 01 33.17 &  +19 16 10.7 &     113688 &      8.670 &      0.090 &      0.610 &      0.024 & 12.07$\pm$0.88 &       5757 &       4.29 &      -0.14 & 2014A\&A...562A..71B \\
       163 & 23 03 57.30 &$-$04 47 42.0 &     113896 &      6.680 &      0.052 &      0.580 &      0.007 & 34.03$\pm$0.77 &       5872 &       4.28 &      -0.18 & 2004AJ....128.1177V \\
       164 & 23 10 43.49 &  +18 54 32.6 &     114450 &      8.560 &      0.055 &      0.595 &      0.043 & 14.48$\pm$1.09 &       5935 &       4.40 &      -0.05 & 2014A\&A...562A..71B \\
       165 & 23 14 07.47 &$-$08 55 27.6 &     114702 &      7.554 &     -0.037 &      0.554 &      0.008 & 25.6$\pm$1.26 &       6028 &       4.33 &      -0.34 & 2014A\&A...562A..71B \\
       166 & 23 16 42.30 &  +53 12 49.0 &     114924 &      5.580 &      0.011 &      0.524 &      0.005 & 48.77$\pm$0.26 &       6134 &       4.21 &       0.00 & 2004AJ....128.1177V \\
       167 & 23 31 31.50 &$-$04 05 15.0 &     116106 &      6.500 &     -0.026 &      0.527 &      0.007 & 38.29$\pm$0.54 &       6005 &       4.42 &      -0.24 & 2004AJ....128.1177V \\
       168 & 23 57 33.52 &$-$09 38 51.1 &     118115 &      7.863 &      0.146 &      0.641 &      0.010 & 20.84$\pm$0.87 &       5833 &       4.39 &       0.02 & 2014A\&A...562A..71B \\
\hline
\end{tabular}
\end{center}
}
\end{table*}

\begin{table*}
\setlength{\tabcolsep}{1.75pt} 
{\scriptsize
\caption{Data used in the calibrations. The columns give: ID,  Hipparcos number, de-reddened apparent magnitude $V_0$ and $(B-V)_0$, $(U-B)_0$ colours, $(U-B)_{Hya}$ colour for the Hyades cluster, UV excess $\delta(U-B)$, normalized factor $f$, normalized UV excess $\delta_{0.6}(U-B)$, original absolute magnitude $(M_V)_{org}$, absolute magnitude for the Hyades cluster $(M_V)_{Hya}$, absolute magnitude offset $\Delta M_V$, estimated absolute magnitude $(M_V)_{est}$, and absolute magnitude residuals $(M_V)_{res}$.} 
\begin{center}
\begin{tabular}{cccccccccccccccc}
\hline
        ID &        Hip &      $V_0$ &  $(B-V)_0$ &  $(U-B)_0$ & $(U-B)_{Hya}$ &   $\delta$ &        $f$ & $\delta_{0.60}$ & $[Fe/H]_{cal}$ & $[Fe/H]_{res}$ & $(M_V)_{org}$ & $(M_V)_{Hya}$ & $\Delta M_V$ & $(M_V)_{est}$ & $(M_V)_{res}$ \\
\hline
         1 &        493 &      7.435 &      0.555 &      0.036 &      0.078 &      0.042 &      1.035 &      0.043 &      -0.12 &       0.17 &      4.586 &      4.212 &      0.374 &      4.260 &      0.326 \\
         2 &        910 &      4.886 &      0.489 &     -0.016 &      0.026 &      0.042 &      1.102 &      0.046 &      -0.14 &       0.19 &      3.521 &      3.751 &     -0.230 &      3.813 &     -0.292 \\
         3 &       1599 &      4.225 &      0.571 &      0.017 &      0.094 &      0.077 &      1.025 &      0.079 &      -0.30 &      -0.11 &      4.556 &      4.318 &      0.238 &      4.542 &      0.014 \\
         4 &       1976 &      7.548 &      0.611 &      0.152 &      0.139 &     -0.013 &      1.015 &     -0.013 &       0.18 &      -0.01 &      4.187 &      4.569 &     -0.382 &      4.325 &     -0.138 \\
         5 &       2711 &      5.563 &      0.464 &     -0.017 &      0.013 &      0.030 &      1.132 &      0.034 &      -0.07 &      -0.13 &      3.532 &      3.570 &     -0.038 &      3.570 &     -0.038 \\
         6 &       3182 &      8.695 &      0.627 &      0.084 &      0.158 &      0.074 &      1.019 &      0.075 &      -0.28 &       0.01 &      4.816 &      4.664 &      0.152 &      4.870 &     -0.054 \\
         7 &       3479 &      7.777 &      0.663 &      0.155 &      0.206 &      0.051 &      1.042 &      0.053 &      -0.17 &       0.09 &      5.226 &      4.867 &      0.359 &      4.963 &      0.263 \\
         8 &       3497 &      6.538 &      0.652 &      0.106 &      0.191 &      0.085 &      1.033 &      0.088 &      -0.35 &      -0.03 &      4.820 &      4.806 &      0.014 &      5.074 &     -0.254 \\
         9 &       3704 &      7.819 &      0.536 &     -0.044 &      0.061 &      0.105 &      1.050 &      0.110 &      -0.47 &      -0.12 &      4.379 &      4.084 &      0.295 &      4.466 &     -0.087 \\
        10 &       3909 &      5.178 &      0.507 &     -0.005 &      0.038 &      0.043 &      1.081 &      0.046 &      -0.14 &      -0.11 &      4.191 &      3.881 &      0.310 &      3.944 &      0.247 \\
        11 &       4544 &      9.630 &      0.492 &     -0.135 &      0.028 &      0.163 &      1.098 &      0.179 &      -0.92 &      -0.05 &      4.456 &      3.773 &      0.683 &      4.538 &     -0.082 \\
        12 &       4892 &      8.501 &      0.575 &      0.004 &      0.098 &      0.094 &      1.023 &      0.096 &      -0.39 &      -0.10 &      4.586 &      4.344 &      0.242 &      4.654 &     -0.068 \\
        13 &       5163 &      9.416 &      0.584 &     -0.007 &      0.108 &      0.115 &      1.019 &      0.117 &      -0.51 &       0.23 &      4.591 &      4.402 &      0.189 &      4.820 &     -0.229 \\
        14 &       5301 &      8.384 &      0.632 &      0.157 &      0.164 &      0.007 &      1.021 &      0.007 &       0.07 &       0.18 &      4.688 &      4.693 &     -0.005 &      4.557 &      0.131 \\
        15 &       5862 &      4.954 &      0.570 &      0.097 &      0.093 &     -0.004 &      1.025 &     -0.004 &       0.13 &      -0.04 &      4.057 &      4.311 &     -0.254 &      4.115 &     -0.058 \\
        16 &       6159 &      8.862 &      0.583 &     -0.027 &      0.107 &      0.134 &      1.019 &      0.137 &      -0.63 &       0.04 &      4.793 &      4.395 &      0.398 &      4.916 &     -0.123 \\
        17 &       7217 &      8.960 &      0.598 &      0.043 &      0.123 &      0.080 &      1.016 &      0.081 &      -0.31 &       0.17 &      4.782 &      4.490 &      0.292 &      4.726 &      0.056 \\
        18 &       7459 &     10.070 &      0.520 &     -0.167 &      0.048 &      0.215 &      1.066 &      0.229 &      -1.36 &      -0.13 &      5.251 &      3.972 &      1.279 &      5.073 &      0.178 \\
        19 &       7978 &      5.531 &      0.527 &     -0.006 &      0.053 &      0.059 &      1.059 &      0.062 &      -0.22 &      -0.27 &      4.324 &      4.021 &      0.303 &      4.163 &      0.161 \\
        20 &       8859 &      6.730 &      0.444 &     -0.062 &      0.006 &      0.068 &      1.156 &      0.079 &      -0.30 &       0.00 &      3.642 &      3.423 &      0.219 &      3.645 &     -0.003 \\
        21 &       9085 &      6.088 &      0.477 &     -0.062 &      0.019 &      0.081 &      1.116 &      0.090 &      -0.36 &      -0.10 &      3.942 &      3.664 &      0.278 &      3.945 &     -0.003 \\
        22 &      10449 &      9.055 &      0.571 &     -0.077 &      0.094 &      0.171 &      1.025 &      0.175 &      -0.89 &      -0.02 &      5.058 &      4.318 &      0.740 &      5.060 &     -0.002 \\
        23 &      10652 &      8.954 &      0.592 &     -0.008 &      0.117 &      0.125 &      1.017 &      0.127 &      -0.57 &       0.10 &      4.965 &      4.452 &      0.513 &      4.922 &      0.043 \\
        24 &      12294 &     10.470 &      0.443 &     -0.171 &      0.006 &      0.177 &      1.157 &      0.205 &      -1.13 &      -0.18 &      4.295 &      3.416 &      0.879 &      4.346 &     -0.051 \\
        25 &      12306 &      7.306 &      0.569 &     -0.015 &      0.092 &      0.107 &      1.026 &      0.110 &      -0.47 &       0.05 &      4.631 &      4.305 &      0.326 &      4.685 &     -0.054 \\
        26 &      12444 &      5.766 &      0.517 &     -0.013 &      0.045 &      0.058 &      1.070 &      0.062 &      -0.22 &      -0.31 &      4.078 &      3.951 &      0.127 &      4.091 &     -0.013 \\
        27 &      12653 &      5.394 &      0.553 &      0.073 &      0.076 &      0.003 &      1.036 &      0.003 &       0.09 &      -0.18 &      4.220 &      4.199 &      0.021 &      4.042 &      0.178 \\
        28 &      12777 &      4.090 &      0.482 &     -0.009 &      0.022 &      0.031 &      1.110 &      0.034 &      -0.08 &      -0.06 &      3.858 &      3.701 &      0.157 &      3.704 &      0.154 \\
        29 &      12889 &      7.625 &      0.576 &      0.043 &      0.099 &      0.056 &      1.022 &      0.057 &      -0.19 &      -0.06 &      4.147 &      4.350 &     -0.203 &      4.466 &     -0.319 \\
        30 &      13366 &      8.234 &      0.499 &     -0.096 &      0.032 &      0.128 &      1.090 &      0.140 &      -0.65 &       0.04 &      4.307 &      3.823 &      0.484 &      4.360 &     -0.053 \\
        31 &      14241 &      8.048 &      0.658 &      0.108 &      0.199 &      0.091 &      1.038 &      0.094 &      -0.38 &       0.07 &      5.325 &      4.840 &      0.485 &      5.142 &      0.183 \\
        32 &      15131 &      6.748 &      0.576 &     -0.023 &      0.099 &      0.122 &      1.022 &      0.125 &      -0.55 &      -0.12 &      4.830 &      4.350 &      0.480 &      4.807 &      0.023 \\
        33 &      15158 &      8.460 &      0.501 &      0.009 &      0.034 &      0.025 &      1.088 &      0.027 &      -0.04 &       0.02 &      3.576 &      3.838 &     -0.262 &      3.804 &     -0.228 \\
        34 &      15381 &      7.472 &      0.598 &      0.154 &      0.123 &     -0.031 &      1.016 &     -0.031 &       0.28 &       0.20 &      3.994 &      4.490 &     -0.496 &      4.144 &     -0.150 \\
        35 &      16169 &      8.210 &      0.612 &      0.022 &      0.140 &      0.118 &      1.015 &      0.120 &      -0.53 &       0.04 &      4.860 &      4.575 &      0.285 &      5.006 &     -0.146 \\
        36 &      17147 &      6.664 &      0.532 &     -0.092 &      0.057 &      0.149 &      1.054 &      0.157 &      -0.76 &       0.05 &      4.626 &      4.056 &      0.570 &      4.691 &     -0.065 \\
        37 &      22632 &      9.106 &      0.481 &     -0.196 &      0.021 &      0.217 &      1.111 &      0.241 &      -1.48 &       0.14 &      4.986 &      3.694 &      1.292 &      4.884 &      0.102 \\
        38 &      23555 &      6.296 &      0.528 &      0.039 &      0.054 &      0.015 &      1.058 &      0.016 &       0.02 &      -0.20 &      3.788 &      4.028 &     -0.240 &      3.937 &     -0.149 \\
        39 &      23688 &      9.647 &      0.416 &     -0.163 &      0.003 &      0.166 &      1.186 &      0.197 &      -1.06 &      -0.17 &      4.367 &      3.220 &      1.147 &      4.097 &      0.270 \\
        40 &      24030 &      9.648 &      0.497 &     -0.147 &      0.031 &      0.178 &      1.092 &      0.194 &      -1.04 &      -0.04 &      4.336 &      3.809 &      0.527 &      4.670 &     -0.334 \\
        41 &      25860 &      8.506 &      0.627 &      0.088 &      0.158 &      0.070 &      1.019 &      0.071 &      -0.26 &       0.09 &      5.019 &      4.664 &      0.355 &      4.850 &      0.169 \\
        42 &      26617 &     10.116 &      0.554 &     -0.025 &      0.077 &      0.102 &      1.035 &      0.106 &      -0.45 &       0.30 &      4.836 &      4.206 &      0.630 &      4.564 &      0.272 \\
        43 &      27072 &      3.587 &      0.480 &     -0.008 &      0.021 &      0.029 &      1.113 &      0.032 &      -0.06 &      -0.04 &      3.833 &      3.686 &      0.147 &      3.678 &      0.155 \\
        44 &      28403 &      8.575 &      0.620 &      0.046 &      0.149 &      0.103 &      1.017 &      0.105 &      -0.44 &      -0.11 &      5.041 &      4.623 &      0.418 &      4.977 &      0.064 \\
        45 &      28671 &      9.260 &      0.602 &     -0.055 &      0.128 &      0.183 &      1.015 &      0.186 &      -0.97 &       0.14 &      5.388 &      4.514 &      0.874 &      5.320 &      0.068 \\
        46 &      29432 &      6.826 &      0.625 &      0.119 &      0.156 &      0.037 &      1.018 &      0.038 &      -0.09 &       0.07 &      4.970 &      4.652 &      0.318 &      4.671 &      0.299 \\
        47 &      31188 &      8.514 &      0.525 &     -0.094 &      0.052 &      0.146 &      1.061 &      0.155 &      -0.75 &      -0.16 &      4.639 &      4.007 &      0.632 &      4.630 &      0.009 \\
        48 &      33582 &      8.987 &      0.569 &     -0.024 &      0.092 &      0.116 &      1.026 &      0.119 &      -0.52 &       0.10 &      4.482 &      4.305 &      0.177 &      4.732 &     -0.250 \\
        49 &      34017 &      5.921 &      0.588 &      0.058 &      0.112 &      0.054 &      1.018 &      0.055 &      -0.18 &      -0.06 &      4.512 &      4.427 &      0.085 &      4.532 &     -0.020 \\
        50 &      34511 &      7.981 &      0.614 &      0.116 &      0.142 &      0.026 &      1.016 &      0.026 &      -0.03 &       0.08 &      4.657 &      4.587 &      0.070 &      4.549 &      0.108 \\
        51 &      35139 &      7.749 &      0.605 &     -0.018 &      0.131 &      0.149 &      1.015 &      0.151 &      -0.72 &      -0.20 &      5.202 &      4.533 &      0.669 &      5.135 &      0.067 \\
        52 &      36491 &      8.482 &      0.514 &     -0.120 &      0.043 &      0.163 &      1.073 &      0.175 &      -0.89 &       0.06 &      5.009 &      3.930 &      1.079 &      4.670 &      0.339 \\
        53 &      36849 &      8.922 &      0.501 &     -0.116 &      0.034 &      0.150 &      1.088 &      0.163 &      -0.80 &       0.04 &      4.384 &      3.838 &      0.546 &      4.508 &     -0.124 \\
        54 &      37419 &      8.655 &      0.616 &      0.064 &      0.145 &      0.081 &      1.016 &      0.082 &      -0.32 &       0.01 &      5.067 &      4.599 &      0.468 &      4.840 &      0.227 \\
        55 &      38541 &      8.288 &      0.615 &     -0.128 &      0.143 &      0.271 &      1.016 &      0.275 &      -1.88 &      -0.04 &      5.964 &      4.593 &      1.371 &      6.066 &     -0.102 \\
        56 &      38769 &      8.780 &      0.504 &     -0.072 &      0.036 &      0.108 &      1.084 &      0.117 &      -0.51 &       0.28 &      4.095 &      3.859 &      0.236 &      4.276 &     -0.181 \\
        57 &      40118 &      6.817 &      0.676 &      0.125 &      0.224 &      0.099 &      1.056 &      0.105 &      -0.44 &       0.03 &      5.126 &      4.936 &      0.190 &      5.289 &     -0.163 \\
        58 &      40778 &      9.699 &      0.467 &     -0.222 &      0.014 &      0.236 &      1.129 &      0.266 &      -1.77 &      -0.22 &      4.774 &      3.592 &      1.182 &      4.987 &     -0.213 \\
        59 &      41484 &      6.312 &      0.623 &      0.118 &      0.153 &      0.035 &      1.017 &      0.036 &      -0.08 &       0.00 &      4.575 &      4.641 &     -0.066 &      4.650 &     -0.075 \\
        60 &      42356 &      7.560 &      0.626 &      0.131 &      0.157 &      0.026 &      1.018 &      0.026 &      -0.03 &      -0.20 &      4.360 &      4.658 &     -0.298 &      4.621 &     -0.261 \\
        61 &      43595 &     10.789 &      0.587 &     -0.049 &      0.111 &      0.160 &      1.018 &      0.163 &      -0.80 &       0.04 &      5.269 &      4.421 &      0.848 &      5.089 &      0.180 \\
        62 &      43726 &      6.007 &      0.665 &      0.211 &      0.209 &     -0.002 &      1.044 &     -0.002 &       0.12 &       0.11 &      4.806 &      4.878 &     -0.072 &      4.693 &      0.113 \\
        63 &      44116 &      8.456 &      0.432 &     -0.109 &      0.004 &      0.113 &      1.169 &      0.132 &      -0.60 &      -0.02 &      3.975 &      3.336 &      0.639 &      3.833 &      0.142 \\
        64 &      44713 &      7.261 &      0.651 &      0.191 &      0.189 &     -0.002 &      1.032 &     -0.002 &       0.12 &      -0.02 &      4.404 &      4.801 &     -0.397 &      4.616 &     -0.212 \\
        65 &      44811 &      7.622 &      0.532 &     -0.063 &      0.057 &      0.120 &      1.054 &      0.126 &      -0.57 &       0.07 &      4.570 &      4.056 &      0.514 &      4.523 &      0.047 \\
        66 &      47048 &      8.451 &      0.366 &     -0.118 &      0.018 &      0.136 &      1.227 &      0.167 &      -0.83 &      -0.33 &      3.507 &      2.875 &      0.632 &      3.567 &     -0.060 \\
        67 &      47174 &      9.940 &      0.623 &      0.025 &      0.153 &      0.128 &      1.017 &      0.130 &      -0.59 &      -0.14 &      5.334 &      4.641 &      0.693 &      5.127 &      0.207 \\
        68 &      48209 &      9.601 &      0.643 &      0.078 &      0.179 &      0.101 &      1.027 &      0.104 &      -0.43 &       0.22 &      5.187 &      4.756 &      0.431 &      5.104 &      0.083 \\
        69 &      49285 &      8.073 &      0.668 &      0.191 &      0.213 &      0.022 &      1.047 &      0.023 &      -0.02 &       0.19 &      4.697 &      4.894 &     -0.197 &      4.839 &     -0.142 \\
        70 &      49615 &      7.703 &      0.504 &     -0.064 &      0.036 &      0.100 &      1.084 &      0.108 &      -0.46 &      -0.03 &      4.340 &      3.859 &      0.481 &      4.231 &      0.109 \\
        71 &      49793 &      8.054 &      0.587 &     -0.026 &      0.111 &      0.137 &      1.018 &      0.139 &      -0.65 &      -0.02 &      4.823 &      4.421 &      0.402 &      4.958 &     -0.135 \\
        72 &      49942 &      8.397 &      0.625 &      0.110 &      0.156 &      0.046 &      1.018 &      0.047 &      -0.14 &       0.13 &      4.531 &      4.652 &     -0.121 &      4.717 &     -0.186 \\
        73 &      50834 &      9.239 &      0.412 &     -0.089 &      0.003 &      0.092 &      1.190 &      0.109 &      -0.47 &      -0.20 &      3.779 &      3.192 &      0.587 &      3.570 &      0.209 \\
        74 &      50965 &      9.753 &      0.555 &     -0.011 &      0.078 &      0.089 &      1.035 &      0.092 &      -0.37 &       0.20 &      4.558 &      4.212 &      0.346 &      4.502 &      0.056 \\
        75 &      52673 &      9.531 &      0.627 &      0.021 &      0.158 &      0.137 &      1.019 &      0.140 &      -0.65 &       0.01 &      5.272 &      4.664 &      0.608 &      5.201 &      0.071 \\
        76 &      52771 &     10.203 &      0.488 &     -0.232 &      0.025 &      0.257 &      1.103 &      0.283 &      -1.98 &       0.00 &      5.299 &      3.744 &      1.555 &      5.290 &      0.009 \\
        77 &      53537 &      7.928 &      0.616 &      0.117 &      0.145 &      0.028 &      1.016 &      0.028 &      -0.04 &      -0.09 &      4.468 &      4.599 &     -0.131 &      4.572 &     -0.104 \\
        78 &      53721 &      5.027 &      0.605 &      0.123 &      0.131 &      0.008 &      1.015 &      0.008 &       0.06 &       0.05 &      4.287 &      4.533 &     -0.246 &      4.402 &     -0.115 \\
        79 &      54469 &      9.749 &      0.542 &     -0.035 &      0.066 &      0.101 &      1.045 &      0.106 &      -0.45 &      -0.02 &      4.584 &      4.125 &      0.459 &      4.483 &      0.101 \\
        80 &      54641 &      8.083 &      0.458 &     -0.176 &      0.011 &      0.187 &      1.139 &      0.213 &      -1.20 &      -0.06 &      4.402 &      3.526 &      0.876 &      4.511 &     -0.109 \\
        81 &      54924 &      9.018 &      0.568 &     -0.008 &      0.091 &      0.099 &      1.026 &      0.102 &      -0.42 &       0.38 &      4.901 &      4.298 &      0.603 &      4.636 &      0.265 \\
        82 &      55592 &      9.942 &      0.470 &     -0.147 &      0.016 &      0.163 &      1.125 &      0.183 &      -0.95 &       0.07 &      4.550 &      3.613 &      0.937 &      4.405 &      0.145 \\
        83 &      55761 &      7.847 &      0.536 &     -0.043 &      0.061 &      0.104 &      1.050 &      0.109 &      -0.47 &       0.15 &      4.651 &      4.084 &      0.567 &      4.461 &      0.190 \\
        84 &      56664 &      8.636 &      0.412 &     -0.109 &      0.003 &      0.112 &      1.190 &      0.133 &      -0.61 &       0.07 &      3.513 &      3.192 &      0.321 &      3.695 &     -0.182 \\
        85 &      57017 &      7.508 &      0.463 &     -0.065 &      0.013 &      0.078 &      1.133 &      0.088 &      -0.35 &       0.08 &      3.876 &      3.562 &      0.314 &      3.833 &      0.043 \\
\hline
\end{tabular}
\end{center}
}
\end{table*}

\begin{table*}
\setcounter{table}{1}
\setlength{\tabcolsep}{1.5pt} 
{\scriptsize
\begin{center}
\begin{tabular}{cccccccccccccccc}
\hline
        ID &        Hip &      $V_0$ &  $(B-V)_0$ &  $(U-B)_0$ & $(U-B)_{Hya}$ &   $\delta$ &        $f$ & $\delta_{0.60}$ & $[Fe/H]_{cal}$ & $[Fe/H]_{res}$ & $(M_V)_{org}$ & $(M_V)_{Hya}$ & $\Delta M_V$ & $(M_V)_{est}$ & $(M_V)_{res}$ \\
\hline
        86 &      57265 &     10.318 &      0.462 &     -0.143 &      0.012 &      0.155 &      1.135 &      0.176 &      -0.90 &       0.03 &      4.297 &      3.555 &      0.742 &      4.301 &     -0.004 \\
        87 &      57443 &      4.879 &      0.662 &      0.098 &      0.204 &      0.106 &      1.042 &      0.110 &      -0.47 &      -0.12 &      5.055 &      4.861 &      0.194 &      5.244 &     -0.189 \\
        88 &      57450 &      9.902 &      0.555 &     -0.154 &      0.078 &      0.232 &      1.035 &      0.240 &      -1.47 &       0.03 &      5.447 &      4.212 &      1.235 &      5.395 &      0.052 \\
        89 &      58843 &      9.194 &      0.579 &     -0.059 &      0.102 &      0.161 &      1.021 &      0.164 &      -0.81 &       0.03 &      4.962 &      4.370 &      0.592 &      5.047 &     -0.085 \\
        90 &      58950 &      7.757 &      0.642 &      0.134 &      0.177 &      0.043 &      1.026 &      0.044 &      -0.12 &       0.07 &      4.981 &      4.750 &      0.231 &      4.801 &      0.180 \\
        91 &      59380 &      7.502 &      0.554 &     -0.059 &      0.077 &      0.136 &      1.035 &      0.141 &      -0.65 &      -0.07 &      4.684 &      4.206 &      0.478 &      4.750 &     -0.066 \\
        92 &      61053 &      6.193 &      0.545 &      0.013 &      0.069 &      0.056 &      1.042 &      0.058 &      -0.20 &      -0.09 &      4.503 &      4.145 &      0.358 &      4.267 &      0.236 \\
        93 &      61619 &      7.338 &      0.476 &     -0.023 &      0.019 &      0.042 &      1.118 &      0.047 &      -0.14 &      -0.18 &      3.591 &      3.657 &     -0.066 &      3.722 &     -0.131 \\
        94 &      62108 &      9.841 &      0.441 &     -0.176 &      0.005 &      0.181 &      1.159 &      0.210 &      -1.17 &       0.36 &      4.402 &      3.402 &      1.000 &      4.365 &      0.037 \\
        95 &      62207 &      5.945 &      0.546 &     -0.045 &      0.070 &      0.115 &      1.042 &      0.120 &      -0.53 &       0.06 &      4.745 &      4.152 &      0.593 &      4.584 &      0.161 \\
        96 &      62809 &      8.424 &      0.596 &     -0.032 &      0.121 &      0.153 &      1.016 &      0.155 &      -0.75 &       0.02 &      5.001 &      4.477 &      0.524 &      5.103 &     -0.102 \\
        97 &      64345 &      8.718 &      0.613 &      0.047 &      0.141 &      0.094 &      1.015 &      0.095 &      -0.39 &       0.21 &      4.855 &      4.581 &      0.274 &      4.887 &     -0.032 \\
        98 &      64394 &      4.228 &      0.570 &      0.067 &      0.093 &      0.026 &      1.025 &      0.027 &      -0.04 &      -0.07 &      4.426 &      4.311 &      0.115 &      4.275 &      0.151 \\
        99 &      64698 &      8.425 &      0.652 &      0.164 &      0.191 &      0.027 &      1.033 &      0.028 &      -0.04 &       0.09 &      4.759 &      4.806 &     -0.047 &      4.776 &     -0.017 \\
       100 &      64747 &      8.271 &      0.637 &      0.118 &      0.171 &      0.053 &      1.023 &      0.054 &      -0.18 &       0.03 &      4.970 &      4.722 &      0.248 &      4.823 &      0.147 \\
       101 &      65418 &     12.138 &      0.433 &     -0.186 &      0.004 &      0.190 &      1.168 &      0.222 &      -1.29 &       0.27 &      4.609 &      3.343 &      1.266 &      4.391 &      0.218 \\
       102 &      66238 &      7.279 &      0.665 &      0.117 &      0.209 &      0.092 &      1.044 &      0.096 &      -0.39 &      -0.17 &      4.889 &      4.878 &      0.011 &      5.188 &     -0.299 \\
       103 &      67655 &      7.941 &      0.653 &     -0.002 &      0.192 &      0.194 &      1.034 &      0.201 &      -1.09 &      -0.16 &      5.920 &      4.812 &      1.108 &      5.714 &      0.206 \\
       104 &      67863 &      9.004 &      0.593 &     -0.055 &      0.118 &      0.173 &      1.016 &      0.176 &      -0.89 &      -0.13 &      5.118 &      4.458 &      0.660 &      5.203 &     -0.085 \\
       105 &      68030 &      6.151 &      0.498 &     -0.087 &      0.032 &      0.119 &      1.091 &      0.130 &      -0.59 &      -0.21 &      4.173 &      3.816 &      0.357 &      4.301 &     -0.128 \\
       106 &      70319 &      6.242 &      0.634 &      0.081 &      0.167 &      0.086 &      1.022 &      0.088 &      -0.35 &       0.06 &      5.065 &      4.705 &      0.360 &      4.974 &      0.091 \\
       107 &      70681 &      9.259 &      0.583 &     -0.113 &      0.107 &      0.220 &      1.019 &      0.224 &      -1.31 &      -0.01 &      5.874 &      4.395 &      1.479 &      5.459 &      0.415 \\
       108 &      70829 &      8.903 &      0.673 &      0.084 &      0.220 &      0.136 &      1.053 &      0.143 &      -0.67 &       0.00 &      5.317 &      4.921 &      0.396 &      5.478 &     -0.161 \\
       109 &      71076 &      7.787 &      0.493 &     -0.085 &      0.028 &      0.113 &      1.097 &      0.124 &      -0.55 &      -0.16 &      3.970 &      3.780 &      0.190 &      4.233 &     -0.263 \\
       110 &      71735 &      7.348 &      0.658 &      0.141 &      0.199 &      0.058 &      1.038 &      0.060 &      -0.21 &       0.14 &      5.230 &      4.840 &      0.390 &      4.971 &      0.259 \\
       111 &      72407 &      9.750 &      0.617 &      0.071 &      0.146 &      0.075 &      1.016 &      0.076 &      -0.29 &       0.25 &      4.887 &      4.605 &      0.282 &      4.815 &      0.072 \\
       112 &      72673 &      7.117 &      0.431 &     -0.117 &      0.003 &      0.120 &      1.171 &      0.141 &      -0.65 &      -0.09 &      3.629 &      3.329 &      0.300 &      3.871 &     -0.242 \\
       113 &      74067 &      7.969 &      0.580 &     -0.057 &      0.103 &      0.160 &      1.020 &      0.163 &      -0.80 &       0.01 &      5.095 &      4.376 &      0.719 &      5.046 &      0.049 \\
       114 &      75181 &      5.639 &      0.635 &      0.066 &      0.168 &      0.102 &      1.022 &      0.104 &      -0.44 &      -0.12 &      4.786 &      4.710 &      0.076 &      5.061 &     -0.275 \\
       115 &      78640 &      9.835 &      0.468 &     -0.202 &      0.015 &      0.217 &      1.127 &      0.245 &      -1.52 &      -0.09 &      4.459 &      3.599 &      0.860 &      4.816 &     -0.357 \\
       116 &      81520 &      7.002 &      0.609 &      0.025 &      0.136 &      0.111 &      1.015 &      0.113 &      -0.49 &      -0.05 &      5.246 &      4.557 &      0.689 &      4.951 &      0.295 \\
       117 &      81681 &      7.160 &      0.604 &      0.050 &      0.130 &      0.080 &      1.015 &      0.081 &      -0.31 &       0.07 &      4.806 &      4.526 &      0.280 &      4.761 &      0.045 \\
       118 &      83229 &      6.987 &      0.568 &     -0.024 &      0.091 &      0.115 &      1.026 &      0.118 &      -0.52 &      -0.03 &      4.564 &      4.298 &      0.266 &      4.720 &     -0.156 \\
       119 &      83489 &      9.078 &      0.635 &      0.088 &      0.168 &      0.080 &      1.022 &      0.082 &      -0.32 &      -0.03 &      4.718 &      4.710 &      0.008 &      4.948 &     -0.230 \\
       120 &      84988 &      6.977 &      0.593 &     -0.016 &      0.118 &      0.134 &      1.016 &      0.136 &      -0.62 &       0.08 &      4.739 &      4.458 &      0.281 &      4.977 &     -0.238 \\
       121 &      85007 &      6.851 &      0.505 &     -0.068 &      0.036 &      0.104 &      1.083 &      0.113 &      -0.48 &      -0.09 &      4.516 &      3.866 &      0.650 &      4.260 &      0.256 \\
       122 &      85042 &      6.217 &      0.659 &      0.220 &      0.200 &     -0.020 &      1.039 &     -0.021 &       0.22 &       0.22 &      4.764 &      4.845 &     -0.081 &      4.559 &      0.205 \\
       123 &      86013 &      8.353 &      0.558 &     -0.059 &      0.081 &      0.140 &      1.032 &      0.144 &      -0.68 &       0.13 &      4.790 &      4.232 &      0.558 &      4.796 &     -0.006 \\
       124 &      86321 &      9.724 &      0.465 &     -0.117 &      0.013 &      0.130 &      1.131 &      0.147 &      -0.69 &       0.18 &      4.340 &      3.577 &      0.763 &      4.155 &      0.185 \\
       125 &      87062 &     10.266 &      0.485 &     -0.209 &      0.024 &      0.233 &      1.107 &      0.258 &      -1.67 &      -0.18 &      5.175 &      3.723 &      1.452 &      5.047 &      0.128 \\
       126 &      88945 &      6.838 &      0.616 &      0.087 &      0.145 &      0.058 &      1.016 &      0.059 &      -0.20 &      -0.22 &      4.864 &      4.599 &      0.265 &      4.724 &      0.140 \\
       127 &      92270 &      6.098 &      0.494 &      0.015 &      0.029 &      0.014 &      1.096 &      0.015 &       0.02 &       0.07 &      3.805 &      3.788 &      0.017 &      3.694 &      0.111 \\
       128 &      92532 &      7.138 &      0.536 &     -0.048 &      0.061 &      0.109 &      1.050 &      0.114 &      -0.50 &       0.06 &      4.712 &      4.084 &      0.628 &      4.488 &      0.224 \\
       129 &      92781 &      9.000 &      0.561 &     -0.067 &      0.084 &      0.151 &      1.030 &      0.156 &      -0.75 &      -0.06 &      4.820 &      4.252 &      0.568 &      4.878 &     -0.058 \\
       130 &      93185 &      6.775 &      0.575 &     -0.004 &      0.098 &      0.102 &      1.023 &      0.104 &      -0.44 &      -0.15 &      4.888 &      4.344 &      0.544 &      4.696 &      0.192 \\
       131 &      94129 &      8.171 &      0.617 &      0.121 &      0.146 &      0.025 &      1.016 &      0.025 &      -0.03 &       0.24 &      4.399 &      4.605 &     -0.206 &      4.562 &     -0.163 \\
       132 &      94645 &      6.232 &      0.533 &      0.065 &      0.058 &     -0.007 &      1.053 &     -0.007 &       0.15 &      -0.09 &      4.032 &      4.063 &     -0.031 &      3.850 &      0.182 \\
       133 &      96124 &      7.129 &      0.664 &      0.150 &      0.207 &      0.057 &      1.043 &      0.059 &      -0.20 &       0.02 &      4.954 &      4.872 &      0.082 &      4.999 &     -0.045 \\
       134 &      96258 &      5.721 &      0.464 &     -0.012 &      0.013 &      0.025 &      1.132 &      0.028 &      -0.04 &      -0.01 &      3.722 &      3.570 &      0.152 &      3.542 &      0.180 \\
       135 &      98355 &      7.445 &      0.468 &     -0.092 &      0.015 &      0.107 &      1.127 &      0.121 &      -0.53 &       0.09 &      3.851 &      3.599 &      0.252 &      4.035 &     -0.184 \\
       136 &      99139 &      8.809 &      0.630 &      0.093 &      0.162 &      0.069 &      1.020 &      0.070 &      -0.26 &       0.11 &      5.079 &      4.682 &      0.397 &      4.863 &      0.216 \\
       137 &      99938 &      8.360 &      0.566 &     -0.031 &      0.089 &      0.120 &      1.027 &      0.123 &      -0.55 &      -0.01 &      4.592 &      4.285 &      0.307 &      4.735 &     -0.143 \\
       138 &     100017 &      5.854 &      0.567 &      0.045 &      0.090 &      0.045 &      1.027 &      0.046 &      -0.14 &       0.05 &      4.630 &      4.292 &      0.338 &      4.354 &      0.276 \\
       139 &     100279 &     10.039 &      0.583 &     -0.070 &      0.107 &      0.177 &      1.019 &      0.180 &      -0.93 &      -0.21 &      5.137 &      4.395 &      0.742 &      5.168 &     -0.031 \\
       140 &     100568 &      8.628 &      0.541 &     -0.139 &      0.065 &      0.204 &      1.046 &      0.213 &      -1.21 &      -0.11 &      5.416 &      4.118 &      1.298 &      5.106 &      0.310 \\
       141 &     100792 &      8.310 &      0.472 &     -0.182 &      0.017 &      0.199 &      1.122 &      0.223 &      -1.30 &      -0.19 &      4.462 &      3.628 &      0.834 &      4.686 &     -0.224 \\
       142 &     102018 &      7.187 &      0.594 &      0.137 &      0.119 &     -0.018 &      1.016 &     -0.018 &       0.21 &       0.08 &      4.167 &      4.465 &     -0.298 &      4.193 &     -0.026 \\
       143 &     102046 &      8.204 &      0.486 &     -0.137 &      0.024 &      0.161 &      1.105 &      0.178 &      -0.91 &       0.14 &      4.245 &      3.730 &      0.515 &      4.488 &     -0.243 \\
       144 &     102762 &      8.070 &      0.575 &      0.067 &      0.098 &      0.031 &      1.023 &      0.032 &      -0.06 &      -0.02 &      4.236 &      4.344 &     -0.108 &      4.333 &     -0.097 \\
       145 &     102805 &      5.990 &      0.414 &     -0.095 &      0.003 &      0.098 &      1.188 &      0.116 &      -0.51 &      -0.19 &      3.560 &      3.206 &      0.354 &      3.620 &     -0.060 \\
       146 &     103458 &      6.507 &      0.582 &     -0.045 &      0.106 &      0.151 &      1.020 &      0.154 &      -0.74 &      -0.13 &      4.781 &      4.389 &      0.392 &      5.007 &     -0.226 \\
       147 &     103498 &      8.271 &      0.514 &     -0.134 &      0.043 &      0.177 &      1.073 &      0.190 &      -1.00 &       0.11 &      4.659 &      3.930 &      0.729 &      4.763 &     -0.104 \\
       148 &     103897 &     10.083 &      0.576 &      0.005 &      0.099 &      0.094 &      1.022 &      0.096 &      -0.39 &       0.28 &      4.515 &      4.350 &      0.165 &      4.660 &     -0.145 \\
       149 &     104659 &      7.343 &      0.502 &     -0.164 &      0.034 &      0.198 &      1.086 &      0.215 &      -1.22 &      -0.14 &      4.662 &      3.845 &      0.817 &      4.844 &     -0.182 \\
       150 &     107294 &      9.993 &      0.462 &     -0.153 &      0.012 &      0.165 &      1.135 &      0.187 &      -0.98 &       0.16 &      4.771 &      3.555 &      1.216 &      4.371 &      0.400 \\
       151 &     107877 &      6.850 &      0.477 &     -0.049 &      0.019 &      0.068 &      1.116 &      0.076 &      -0.29 &      -0.05 &      3.832 &      3.664 &      0.168 &      3.873 &     -0.041 \\
       152 &     108468 &      7.188 &      0.618 &      0.110 &      0.147 &      0.037 &      1.016 &      0.038 &      -0.09 &      -0.01 &      4.569 &      4.611 &     -0.042 &      4.630 &     -0.061 \\
       153 &     108736 &      7.088 &      0.565 &      0.022 &      0.088 &      0.066 &      1.028 &      0.068 &      -0.25 &       0.04 &      4.320 &      4.279 &      0.041 &      4.448 &     -0.128 \\
       154 &     109144 &      7.216 &      0.515 &      0.014 &      0.044 &      0.030 &      1.072 &      0.032 &      -0.06 &       0.02 &      3.668 &      3.937 &     -0.269 &      3.928 &     -0.260 \\
       155 &     109381 &      7.828 &      0.657 &      0.208 &      0.198 &     -0.010 &      1.037 &     -0.010 &       0.16 &       0.03 &      4.686 &      4.834 &     -0.148 &      4.605 &      0.081 \\
       156 &     109646 &      7.408 &      0.510 &     -0.087 &      0.040 &      0.127 &      1.077 &      0.137 &      -0.63 &       0.01 &      4.616 &      3.902 &      0.714 &      4.424 &      0.192 \\
       157 &     110035 &      7.010 &      0.590 &      0.069 &      0.114 &      0.045 &      1.017 &      0.046 &      -0.13 &       0.01 &      4.551 &      4.440 &      0.111 &      4.499 &      0.052 \\
       158 &     110341 &      6.147 &      0.440 &     -0.059 &      0.005 &      0.064 &      1.161 &      0.074 &      -0.28 &      -0.19 &      3.695 &      3.394 &      0.301 &      3.595 &      0.100 \\
       159 &     110560 &     10.544 &      0.540 &     -0.013 &      0.064 &      0.077 &      1.047 &      0.081 &      -0.31 &       0.22 &      4.174 &      4.111 &      0.063 &      4.343 &     -0.169 \\
       160 &     111565 &      7.571 &      0.662 &      0.098 &      0.204 &      0.106 &      1.042 &      0.110 &      -0.47 &       0.01 &      5.085 &      4.861 &      0.224 &      5.244 &     -0.159 \\
       161 &     112811 &      9.283 &      0.664 &      0.069 &      0.207 &      0.138 &      1.043 &      0.144 &      -0.67 &       0.03 &      5.378 &      4.872 &      0.506 &      5.433 &     -0.055 \\
       162 &     113688 &      8.596 &      0.586 &      0.073 &      0.110 &      0.037 &      1.018 &      0.038 &      -0.09 &       0.05 &      4.005 &      4.414 &     -0.409 &      4.433 &     -0.428 \\
       163 &     113896 &      6.658 &      0.573 &      0.047 &      0.096 &      0.049 &      1.024 &      0.050 &      -0.16 &       0.02 &      4.317 &      4.331 &     -0.014 &      4.412 &     -0.095 \\
       164 &     114450 &      8.427 &      0.552 &      0.024 &      0.075 &      0.051 &      1.037 &      0.053 &      -0.17 &      -0.12 &      4.231 &      4.192 &      0.039 &      4.287 &     -0.056 \\
       165 &     114702 &      7.529 &      0.546 &     -0.043 &      0.070 &      0.113 &      1.042 &      0.118 &      -0.51 &      -0.17 &      4.570 &      4.152 &      0.418 &      4.573 &     -0.003 \\
       166 &     114924 &      5.564 &      0.519 &      0.007 &      0.047 &      0.040 &      1.067 &      0.043 &      -0.12 &      -0.12 &      4.005 &      3.965 &      0.040 &      4.009 &     -0.004 \\
       167 &     116106 &      6.480 &      0.520 &     -0.031 &      0.048 &      0.079 &      1.066 &      0.084 &      -0.33 &      -0.09 &      4.395 &      3.972 &      0.423 &      4.222 &      0.173 \\
       168 &     118115 &      7.832 &      0.631 &      0.139 &      0.163 &      0.024 &      1.020 &      0.024 &      -0.02 &      -0.04 &      4.426 &      4.687 &     -0.261 &      4.640 &     -0.214 \\
\hline
\end{tabular}
\end{center}
}
\end{table*}

\end{document}